\documentclass[11pt,a4paper]{article}

\usepackage[a4paper,margin=27mm,headheight=23pt,headsep=7mm,footskip=13mm]{geometry}
\usepackage[T1]{fontenc}
\usepackage{mlmodern}
\usepackage{amsmath,amssymb,mathtools,bm,amsthm}
\usepackage{microtype}
\usepackage{xcolor}
\usepackage{enumitem}
\usepackage{titlesec}
\usepackage{fancyhdr}
\usepackage{hyperref}
\usepackage[nameinlink,capitalise]{cleveref}
\usepackage[numbers,sort&compress]{natbib}
\usepackage[style=compact-left]{tdqtitle}
\usepackage{comment}

\numberwithin{equation}{section}

\definecolor{TextInk}{HTML}{24313F}
\definecolor{FormalBlue}{HTML}{3C5F74}
\definecolor{InsightTeal}{HTML}{5C7E77}
\definecolor{NeutralGray}{HTML}{6D7479}
\definecolor{RuleGray}{HTML}{D5D9DA}

\definecolor{darkred}{rgb}{0.7, 0., 0.}
\definecolor{orangered}{rgb}{1,0.27,0.}
\definecolor{steelblue}{rgb}{0.275,0.51, 0.706}
\definecolor{royalblue}{rgb}{0.2549,0.4118,0.8824}
\definecolor{forestgreen}{rgb}{0.13,0.55,0.13}
\definecolor{rossoferrari}{HTML}{D9073D}
\definecolor{mediumblue}{HTML}{0000CD}

\newcommand{\TC}{\mathcal{T}_{C}}
\newcommand{\TCstar}{\mathcal{T}_{C}^{\star}}

\TDQTitleSize{26}{32}
\TDQSubtitleSize{18}{23}

\TDQTitle
  [
  In-in formalism \& $\boldsymbol{\TCstar}$-product:
  ]
  {
  In-in formalism and \texorpdfstring{$\TCstar$}{TC-star}-product:
  no contribution from total time derivatives
  }
\TDQSubtitle{no contribution from total time derivatives}

\TDQPreprint{FERMILAB-PUB-26-0816-T}
\TDQPreprint{RIKEN-iTHEMS-Report-26}
\TDQPreprint{KOBE-COSMO-26-14}
\TDQPreprint{KEK-TH-2886}

\TDQAuthor{Yohei Ema}{a}
\TDQAuthor{Muzi Hong}{b,c}
\TDQAuthor{Ryusuke Jinno}{d}
\TDQAuthor{Kyohei Mukaida}{e}

\TDQAffiliation{a}
  {University of Florida}
  {Institute for Fundamental Theory, Department of Physics,
   Gainesville, FL 32611, USA}
\TDQAffiliation{b}
{Fermi National Accelerator Laboratory}
{Batavia, IL 60510, USA}
\TDQAffiliation{c}
  {RIKEN iTHEMS}
  {Center for Interdisciplinary Theoretical and Mathematical Sciences,
   Wako, Saitama 351-0198, Japan}
\TDQAffiliation{d}
  {Kobe University}
  {Department of Physics, Graduate School of Science,
   1-1 Rokkodai, Kobe, Hyogo 657-8501, Japan}
\TDQAffiliation{e}
  {KEK Theory Center / SOKENDAI}
  {IPNS, KEK; Graduate University for Advanced Studies,
   1-1 Oho, Tsukuba, Ibaraki 305-0801, Japan}

\TDQAbstract{
In the in-in formalism, total time derivatives are sometimes argued to contribute nontrivially to inflationary correlators at the evaluation point, \emph{i.e.}, at the end of inflation.
In this paper, we disprove this claim based on the path integral.
After formulating the canonical quantization and the regulated path integral of systems with total time derivatives, we derive the $\TCstar$-ordering, the correct operator ordering on the closed time path.
In the $\TCstar$-ordering, the evaluation point, or the turning point of the in-in contour, is merely an intermediate point and is never an endpoint of the generalized path integral contour.
As a result, any apparent boundary contributions at the turning point of the forward and backward branches completely cancel out.
We demonstrate this cancellation by re-analyzing examples in the literature, with an emphasis on contact terms associated with the $\TCstar$-ordering overlooked previously.
}

\hypersetup{
  colorlinks=true,
  linkcolor=mediumblue,
  citecolor=mediumblue,
  urlcolor=mediumblue,
  linktoc=all,
  bookmarksopen=true,
  bookmarksnumbered=true,
  pdfsubject={Canonical quantization and regulated path integrals for total derivatives},
  pdfkeywords={total derivative, Dirac bracket, path integral, T-star product, closed time path}
}

\color{black}
\setlist{
  topsep=4pt,
  itemsep=2pt,
  parsep=0pt,
  leftmargin=2.1em
}
\allowdisplaybreaks[2]
\titleformat{\section}
  {\normalfont\Large\bfseries}
  {\thesection}{1em}{}
\titlespacing*{\section}
  {0pt}
  {3.5ex plus 1ex minus 0.2ex}
  {2.3ex plus 0.2ex}

\titleformat{\subsection}
  {\normalfont\large\bfseries}
  {\thesubsection}{1em}{}
\titlespacing*{\subsection}
  {0pt}
  {3.25ex plus 1ex minus 0.2ex}
  {1.5ex plus 0.2ex}

\titleformat{\subsubsection}
  {\normalfont\normalsize\bfseries}
  {\thesubsubsection}{1em}{}
\titlespacing*{\subsubsection}
  {0pt}
  {3.25ex plus 1ex minus 0.2ex}
  {1.5ex plus 0.2ex}

\newcommand{\dd}{\mathrm{d}}
\newcommand{\ii}{i}

\newcommand{\Tstar}{\mathcal{T}^{\star}}

\newtheoremstyle{tdqplain}
  {6pt}{6pt}{\itshape}{}{\bfseries\color{FormalBlue}}{.}{0.5em}{}
\theoremstyle{tdqplain}

\newtheoremstyle{tdqdef}
  {6pt}{6pt}{\normalfont}{}{\bfseries\color{FormalBlue}}{.}{0.5em}{}
\theoremstyle{tdqdef}

\crefname{theorem}{theorem}{theorems}
\crefname{lemma}{lemma}{lemmas}
\crefname{proposition}{proposition}{propositions}
\crefname{definition}{definition}{definitions}
\crefname{remark}{remark}{remarks}

\begin{document}
\maketdqtitle

\tableofcontents
\newpage

\section{Introduction}
\label{sec:introduction}

In quantum field theory, it is often the case that a given theory is not represented by a single, unique action.
Its apparent form changes by integration by parts and field redefinitions.
The fact that physical observables remain unchanged under such rewriting is a fundamental premise,
and it is extensively used in, \textit{e.g.}, effective field theory calculations~\cite{Arzt:1993gz,Burgess:2007pt,Brivio:2017vri,Manohar:2018aog}.
One particular example of the redundancy is total derivative terms, the addition of which does not affect observables as long as the boundary conditions are kept fixed.

This property is relevant in real-time quantum field theory, especially in the in-in formalism widely used for inflationary  correlators~\cite{Maldacena:2002vr,Weinberg:2005vy}.
In such a calculation, the interaction Lagrangian is often reorganized by integration by parts, generating total time derivatives~\cite{Maldacena:2002vr}.
It has been recently argued that when the quantity inside a total time derivative contains time derivatives of fields, a nonzero contribution can remain at the evaluation time ({\textit{e.g.},}~\cite{Braglia:2024zsl,Wang:2025qfh,Ma:2026gdh}).
If correct, this claim would challenge the standard lore that actions differing by total time derivatives are physically equivalent in real-time quantum theory,
with implications far beyond the context of cosmological calculations.

The apparent origin of this argument is the split of the closed-time path into forward and backward branches.
When each branch is expressed as a real-time integral, the observation time is an endpoint of each branch, and integrating a total time derivative over each branch produces a boundary term.
When the quantity inside the total derivative depends solely on the field itself, the resulting boundary terms trivially cancel because the fields from the two branches are ``sewn together'' at the observation time; by contrast, if that quantity also depends on the time derivatives of the field, this cancellation seems no longer manifest.

However, the in-in formalism is originally defined along a single smooth closed time path, rather than as two independent real-time evolutions.\footnote{
    In general, the Schwinger-Keldysh formalism is defined on a general smooth path integral contour~$C$, and the in-in contour is a particular choice of the contour $C$~(see~\cite{Landsman:1986uw,Berges:2004yj,Haehl:2016pec,Liu:2018kfw,Haehl:2024pqu} for reviews).
}
Any terms that appear at the turning point by splitting the contour into two branches must therefore cancel out.
This point becomes immediately clear by considering an analogous situation in the in-out formalism.
As far as a total time derivative is concerned, the only difference between the in-in and in-out formalisms is the location of their true endpoints.
In the in-out formalism, the total time derivatives contribute, if at all, only through the endpoints $t = \pm \infty$. 
We are free to split the in-out contour at any intermediate time, possibly at the point of operator insertion for correlation functions, but such an artificial split cannot introduce any additional boundary contributions to physical observables.
For exactly the same reason, the turning point of the in-in contour cannot produce any additional boundary contributions to the correlators.

In the rest of this paper, we will formulate and confirm this contour-level observation.
We will see that the path integral of the in-in formalism results in the $\TCstar$-product, a specific operator ordering along the in-in contour analogous to the $\mathcal{T}^\star$-product in the in-out formalism, which guarantees the absence of contributions at the evaluation point. 
The $\TCstar$-product introduces contact terms not included in the previous studies, and including these terms restores the cancellation of total time derivatives.

\subsection*{Structure of the paper}

To formulate the above argument precisely, we will clarify several points throughout this paper.
First, we consider a quantum mechanical system of the form
\begin{align}
L
&=
L_0 + \frac{\mathrm{d}}{\mathrm{d}t} G (q, \dot{q}, t).
\end{align}
The $\dot{q}$ dependence of $G$ introduces $\ddot{q}$ in the Lagrangian, and this appears to define a higher-derivative system.
In Sec.~\ref{sec:canonical-quantization}, despite its appearance, we show that second-class constraints remove extra degrees of freedom associated with the higher-derivative term, and the system reduces to the one described solely by $L_0$.

Next, we clarify how the equivalence at the level of canonical quantization is reflected in the path integral formalism.
In Sec.~\ref{sec:path-integral-tstar}, we begin with the in-out formalism and derive a discretized path integral based on the canonical quantization derived in Sec.~\ref{sec:canonical-quantization}.
Since the path integral includes configurations $q(t)$ that are not necessarily differentiable, we introduce a regulated definition of $\mathrm{d}G/\mathrm{d}t$.
We express $\dd G/\dd t$ as a finite difference of $G$ in the discretized path integral, and show that its sum vanishes identically.
In addition, we confirm the textbook-level property that the operator ordering associated with time-derivative insertions in the in-out path integral is the $\Tstar$-product rather than the naive $\mathcal{T}$-product.\footnote{
    The notations differ among textbooks. In this paper, we use the terminology in~\cite{Gross:1969zow,Kugo:1997gva,Bastianelli:2006rx} where the $\mathcal{T}^\star$-product indicates time derivatives acting outside of the time-ordering theta functions, whereas the time derivatives in the naive $\mathcal{T}$-product act only on fields and not on the theta functions.
    \cite{Itzykson:1980rh} uses $\hat{\mathcal{T}}$ to indicate our $\mathcal{T}^\star$-product, while \cite{Weinberg:1995mt,Srednicki:2007qs} simply use $\mathcal{T}$ for our $\mathcal{T}^\star$.
    Despite the notational differences, the conclusion is unique: the path integral results in the time ordering where the time derivatives act outside of the theta functions.
    \label{foot:Tstar}
}
In the $\Tstar$-product, the time derivatives act outside of the time-ordering and hence also on the theta functions, which guarantees the absence of the total time derivative contribution in the in-out formalism.

In Sec.~\ref{sec:closed-time-path}, we extend the construction to the closed time path and derive the contour-ordered $\TCstar$-product.
The $\TCstar$-product specifies the treatment of the turning point, introducing contact terms when time derivatives act on the contour-ordering step functions.
Once the contact terms are included, the contributions from the two branches cancel out at the turning point.
Although we work with a quantum mechanical system in Secs.~\ref{sec:canonical-quantization}\,--\,\ref{sec:closed-time-path} for notational simplicity, its extension to quantum field theory is straightforward.

In Sec.~\ref{sec:examples}, we illustrate the cancellation explicitly in examples of inflationary correlators studied in the previous literature, with an emphasis on the contact terms overlooked previously. 
We summarize our results and conclude in Sec.~\ref{sec:discussion_and_conclusions}.

\section{Canonical quantization with total time derivatives}
\label{sec:canonical-quantization}

In this section, we review the canonical quantization of a Lagrangian that contains a total time derivative term.
We will see that it leads to a theory equivalent to the one without the total time derivative, following the Dirac method of quantization in the presence of constraints~\cite{Dirac1964,HenneauxTeitelboim1992}.
We thus consider the Lagrangian
\begin{align}
L
&=
L_0 + \frac{\dd G}{\dd t},
\quad
L_0(q, \dot{q})
=
\frac{M}{2} \dot{q}^{\,2} - V (q),
\label{eq:total-derivative-lagrangian}
\end{align}
where $G = G(q, \dot{q}, t)$, and compare it with the theory governed by $L_0$.
In this paper, we assume that $G = 0$ near $t = t_\mathrm{i}$ and $t_\mathrm{f}$.\footnote{For simplicity, we impose $G|_{t_\mathrm{i}} = G|_{t_\mathrm{f}} = 0$ explicitly. Under the standard assumptions for vacuum projection, correlation functions become independent of this choice in the asymptotic-time limit with the $i\epsilon$-prescription.}

\subsection{First-order formulation and constraint analysis}

Since $G(q,\dot q,t)$ depends on $\dot q$, the total time derivative $\dd G/\dd t$ contains $\ddot q$.
To formulate the theory in first-order form, we introduce two independent variables
\begin{align}
Q_0
&\equiv
q,
\qquad
Q_1
\equiv \dot q.
\end{align}
Introducing a Lagrange multiplier $\lambda$ and writing $G = G(Q_0,Q_1,t)$, the Lagrangian becomes
\begin{align}
L
&=
\frac{M}{2} Q_1^2 - V(Q_0) + G_{Q_0} Q_1 + G_{Q_1} \dot{Q}_1 + G_t + \lambda (\dot{Q}_0 - Q_1),
\label{eq:canonical-higher-L}
\end{align}
where the subscript on $G$ denotes the partial derivative with respect to the corresponding variable.
The equation of motion for $\lambda$ imposes $\dot{Q}_0 = Q_1$, and substituting this relation back into the Lagrangian reproduces the original theory.
The conjugate momenta are
\begin{align}
\Pi_0
&=
\frac{\partial L}{\partial \dot{Q}_0}
=
\lambda,
\qquad
\Pi_1
=
\frac{\partial L}{\partial \dot{Q}_1}
=
G_{Q_1},
\qquad
\Pi_\lambda
=
\frac{\partial L}{\partial \dot{\lambda}}
=
0.
\end{align}
These definitions of momenta do not determine $\dot{Q}_0$, $\dot{Q}_1$, and $\dot{\lambda}$, and therefore imply three primary constraints
\begin{align}
\phi_0
&\equiv
\Pi_0 - \lambda
\simeq
0,
\qquad
\phi_1
\equiv
\Pi_1 - G_{Q_1}
\simeq
0,
\qquad
\phi_\lambda
\equiv
\Pi_\lambda
\simeq
0,
\end{align}
where ``$\simeq$'' indicates the equality under the constraints $\phi_a = 0$.
The Legendre transform gives
\begin{align}
H_T
&=
\Pi_0 \dot{Q}_0 + \Pi_1 \dot{Q}_1 + \Pi_\lambda \dot{\lambda} - L
=
\phi_0 \dot{Q}_0 + \phi_1 \dot{Q}_1 + \phi_\lambda \dot{\lambda} + H_c,
\end{align}
with the constrained Hamiltonian
\begin{align}
H_c
&=
- \frac{M}{2} Q_1^2 + V (Q_0) - G_t + (\lambda - G_{Q_0}) Q_1.
\end{align} 
To make the role of the undetermined velocities as constraint multipliers explicit, we introduce $u_0$, $u_1$, and $u_\lambda$ and write
\begin{align}
H_T
&=
\phi_0 u_0 + \phi_1 u_1 + \phi_\lambda u_\lambda + H_c.
\end{align}

We now apply the Dirac--Bergmann algorithm.
We require the primary constraints to be preserved under time evolution
\begin{align}
\dot{\phi}_a
&=
\frac{\partial \phi_a}{\partial t} + \{ \phi_a, H_T \}
\simeq
\frac{\partial \phi_a}{\partial t} + \{ \phi_a, H_c \} + \sum_{b = 0, 1, \lambda} \{ \phi_a, \phi_b \} u_b
\simeq
0,
\label{eq:phidot}
\end{align}
where $\{f, g\}$ denotes the Poisson bracket defined in the full six-dimensional phase space.
Explicitly, these equations give
\begin{align}
\dot{\phi}_0
&=
- u_\lambda - V' (Q_0) + G_{Q_0 Q_0} Q_1 + G_{Q_0 Q_1} u_1 + G_{Q_0 t},
\\
\dot{\phi}_1
&=
M Q_1 + G_{Q_0} + G_{Q_0 Q_1} Q_1 - G_{Q_0 Q_1} u_0 - \lambda,
\\
\dot{\phi}_\lambda
&=
u_0 - Q_1.
\end{align}
To examine whether these equations determine all the multipliers $u_a$, we consider the matrix $C^{(3)}_{a b} = \{ \phi_a,\phi_b \}$ ($a,b = 0,1,\lambda$), which in the present case takes the form
\begin{align}
C^{(3)}
&=
\begin{pmatrix}
0 & G_{Q_0 Q_1} & -1 \\
-G_{Q_0 Q_1} & 0 & 0 \\
1 & 0 & 0
\end{pmatrix},
\qquad
\det C^{(3)}
=
0.
\end{align}
Since $C^{(3)}$ is singular, the conditions $\dot{\phi}_a \simeq 0$ admit a nontrivial combination in which the multipliers drop out.
Indeed, a null vector of $C^{(3)}$ is found to be
\begin{align}
v^a
&=
\left( 0, \, 1, \, G_{Q_0 Q_1} \right),
\qquad
v^a C^{(3)}_{a b}
=
0,
\end{align}
and thus contracting $v^a$ with Eq.~(\ref{eq:phidot}), we find a secondary constraint
\begin{align}
\phi_2
\equiv
\sum_a
v^a
\left(
\frac{\partial \phi_a}{\partial t}
+ \{ \phi_a,H_c \}
\right)
&=
M Q_1 + G_{Q_0} - \lambda
\simeq
0.
\end{align}
Requiring the secondary constraint to be preserved under time evolution gives
\begin{align}
\dot{\phi}_2
&=
\frac{\partial \phi_2}{\partial t} + \{ \phi_2, H_T \}
\simeq
\frac{\partial \phi_2}{\partial t} + \{ \phi_2, H_c \} + \sum_{b = 0, 1, \lambda} \{ \phi_2, \phi_b \} u_b
\nonumber \\
&=
G_{Q_0 t}
+ G_{Q_0 Q_0} u_0
+ (M + G_{Q_0 Q_1}) u_1
- u_\lambda
\simeq
0.
\label{eq:dphi_2}
\end{align}
Eqs.~(\ref{eq:phidot}) ($a = 0, 1, \lambda$) and (\ref{eq:dphi_2}) determine all three multipliers
\begin{align}
u_0
&=
Q_1,
\qquad
u_1
=
- \frac{V' (Q_0)}{M},
\qquad
u_\lambda
=
- \left( 1 + \frac{G_{Q_0 Q_1}}{M} \right) V' (Q_0) + G_{Q_0 Q_0} Q_1 + G_{Q_0 t},
\end{align}
leaving the already known constraint $\phi_2 \simeq 0$.
Thus no further constraints arise.
The matrix of Poisson brackets $C^{(4)}_{a b} = \{ \phi_a, \phi_b \}$ ($a, b = 0, 1, 2, \lambda$) is given by
\begin{align}
C^{(4)}
&=
\begin{pmatrix}
0 & G_{Q_0 Q_1} & -G_{Q_0 Q_0} & -1 \\
-G_{Q_0 Q_1} & 0 & -(M + G_{Q_0 Q_1}) & 0 \\
G_{Q_0 Q_0} & M + G_{Q_0 Q_1} & 0 & -1 \\
1 & 0 & 1 & 0
\end{pmatrix},
\qquad
\det C^{(4)}
=
M^2,
\end{align}
and the nonzero determinant implies that all four constraints are second class.

\subsection{Reduced phase space and action}

Having identified the four second-class constraints, we now restrict the symplectic potential and the Hamiltonian to the constraint surface. 
In the original six-dimensional phase space, the former is
\begin{align}
\Theta
&\equiv
\Pi_0 \dd Q_0 + \Pi_1 \dd Q_1 + \Pi_\lambda \dd \lambda.
\end{align}
To describe the symplectic structure, we introduce the exterior derivative on the six-dimensional phase space at fixed time as
\begin{align}
\dd_{\rm ph}
&=
\dd Q_0 \frac{\partial}{\partial Q_0}
+ \dd Q_1 \frac{\partial}{\partial Q_1}
+ \dd\lambda \frac{\partial}{\partial \lambda}
+ \dd\Pi_0 \frac{\partial}{\partial \Pi_0}
+ \dd\Pi_1 \frac{\partial}{\partial \Pi_1}
+ \dd\Pi_\lambda \frac{\partial}{\partial \Pi_\lambda}.
\end{align}
Restricting to the constraint surface gives
\begin{align}
\Theta_{\rm red}
&\equiv
\Theta |_{\phi_a = 0}
=
\lambda \dd Q_0 + G_{Q_1} \dd Q_1
=
M Q_1 \dd Q_0 + \dd_{\rm ph} G,
\end{align}
where we used $\dd_{\rm ph} G = G_{Q_0} \dd Q_0 + G_{Q_1} \dd Q_1$ and $\phi_2 = 0$ in the last equality.
The associated symplectic two-form is given by
\begin{align}
\dd_{\rm ph} \Theta_{\rm red}
&=
M \dd Q_1 \wedge \dd Q_0.
\end{align}
It takes the standard expression $\dd_{\rm ph} \Theta_{\rm red} = \dd p \wedge \dd q$ by defining
\begin{align}
q
&= Q_0,
\qquad
p
= M Q_1.
\end{align}
Thus the number of physical degrees of freedom does not change despite the higher derivatives.
In terms of these canonical variables, the symplectic potential reads
\begin{align}
\Theta_{\rm red}
&=
p \dd q + \dd_{\rm ph} G
= p \dd q + G_q \dd q + G_p \dd p.
\end{align}
Restricting the Hamiltonian to the constraint surface, we obtain the reduced Hamiltonian
\begin{align}
H_{\rm red}
&=
\left.H_T\right|_{\phi_a = 0}
=
\left.H_c\right|_{\phi_a = 0}
=
H_0 - G_t,
\qquad
H_0
\equiv
\frac{p^2}{2M} + V(q).
\end{align}
Using $\Theta_\text{red}$,
the reduced phase-space action is therefore
\begin{align}
S_{\rm red}
&=
\int_{t_\mathrm{i}}^{t_\mathrm{f}}
\left[
\Theta_{\rm red} - H_{\rm red}\,\dd t
\right]
=
\int_{t_\mathrm{i}}^{t_\mathrm{f}} \dd t\,
\left[
(p \dot q + G_q \dot q + G_p \dot p) - (H_0 - G_t)
\right]
\nonumber\\
&=
\int_{t_\mathrm{i}}^{t_\mathrm{f}} \dd t\,
\left[
p \dot q
- H_0
\right]
+ G|_{t_\mathrm{f}} - G|_{t_\mathrm{i}}.
\end{align}
Thus the only effect of $G$ on $S_\text{red}$ is the boundary term $G|_{t_\mathrm{f}} - G|_{t_\mathrm{i}}$.\footnote{Hamilton's equations are the same as those governed by $H_0$ without the total derivative, despite the fact that $H_{\rm red}$ takes a different form. See Appendix~\ref{app:Hamiltonian_evolution}.}
Quantizing this reduced system in the standard way, we obtain
\begin{align}
[\hat q,\hat p]= i,
\qquad
\hat{H}
=
\hat{H}_0
=
\frac{\hat{p}^{\,2}}{2 M} + V (\hat{q}).
\label{eq:canonical-physical-H}
\end{align}
Since $G|_{t_\mathrm{f}} = G|_{t_\mathrm{i}} = 0$ by assumption, the time evolution coincides with the theory without the total derivative.

\section[
Path integral and \texorpdfstring{$\Tstar$}{T-star}-product
]{%
Path integral and \texorpdfstring{$\Tstar$}{T-star}-product
}
\label{sec:path-integral-tstar}

In this section, starting from the quantized theory in Sec.~\ref{sec:canonical-quantization}, we construct a discretized path integral, working in the in-out formalism as a warmup.
Although the system evolves with $\hat{H}_0$ as we saw in Sec.~\ref{sec:canonical-quantization}, we can add a sum that vanishes identically to recover the total derivative and make its connection to the calculations in the literature explicit. 
We will see that the discretized prescription of the path integral results in the $\Tstar$-product in the continuum limit, guaranteeing the absence of the total time derivative contribution.

\subsection{Total time derivative in path integral}
\label{sec:path-integral-canonical}

As we have seen in Sec.~\ref{sec:canonical-quantization}, even starting from the Lagrangian with a total time derivative, the quantized theory is governed by $\hat{H}_0$ without this term.
The transition kernel is then given by
\begin{align}
  K(q_\mathrm{f},t_\mathrm{f};q_\mathrm{i},t_\mathrm{i})
  =\langle q_\mathrm{f}|
  \hat{U}_0(t_\mathrm{f},t_\mathrm{i})
  |q_\mathrm{i}\rangle,
  \quad
  \hat{U}_0(t_\mathrm{f},t_\mathrm{i}) = 
  e^{-i\hat H_0(t_\mathrm{f}-t_\mathrm{i})}.
  \label{eq:path-integral-transition-kernel}
\end{align}
To derive the path integral expression of $K$, we divide the interval into $N$ steps as
\begin{align}
  t_n=t_\mathrm{i}+n\epsilon,\quad
  \epsilon=\frac{t_\mathrm{f}-t_\mathrm{i}}{N},\quad
  q_0=q_\mathrm{i},\quad q_N=q_\mathrm{f}.
  \label{eq:path-integral-time-discretization}
\end{align}
Following the standard procedure, we obtain the finite-$N$ path integral expression as
\begin{align}
  K_{N}\equiv
  \int\prod_{n=1}^{N-1}\dd q_n
  \prod_{n=0}^{N-1}\frac{\dd p_{n+\frac12}}{2\pi}
  \exp\!\left[
    i\sum_{n=0}^{N-1}
    \left\{p_{n+\frac12}\Delta q_{n+\frac12}
      -\epsilon\left(\frac{p_{n+\frac12}^2}{2M}+V(\bar q_{n+\frac12})\right)\right\}
  \right],
  \label{eq:path-integral-phase-space-regulator}
\end{align}
where $\Delta q_{n+\frac12}=q_{n+1}-q_n$, and we use the symmetric prescription $\bar q_{n+\frac12}= (q_{n+1}+q_n)/2$ for midpoints.
After performing the momentum integrals explicitly, we arrive at the standard result
\begin{align}
  K_{N}
  &=\left(\frac{M}{2\pi i\epsilon}\right)^{N/2}
    \int\prod_{n=1}^{N-1}\dd q_n\,
    e^{iS_{0,N}[q]},
    \quad
  S_{0,N}[q]
  \equiv\sum_{n=0}^{N-1}\epsilon
    \left[\frac{M}{2}\left(\frac{\Delta q_{n+\frac12}}{\epsilon}\right)^2
      -V(\bar q_{n+\frac12})\right].
  \label{eq:path-integral-discrete-action}
\end{align}
The continuum limit $\epsilon \to 0$ reproduces $K$.
Since the system evolves with $\hat{H}_0$, no contribution from the total time derivative remains in this formula.

Nevertheless, we can make it resemble the original Lagrangian $L = L_0 + \dd G/\dd t$ as follows. 
We define the link-centered quantity $G_{n+\frac{1}{2}}$ and its site-centered finite difference by
\begin{align}
  G_{n+\frac12}
  &\equiv G\!\left(\bar q_{n+\frac12},v_{n+\frac12},t_{n+\frac12}\right),
  \quad
  (\delta_\epsilon G)_n
  \equiv\frac{G_{n+\frac12}-G_{n-\frac12}}{\epsilon},
  \label{eq:discrete-total-derivative}
\end{align}
where $v_{n+\frac12}\equiv\Delta q_{n+\frac12}/\epsilon = (q_{n+1}-q_n)/\epsilon$ and $t_{n+\frac12}\equiv(t_{n+1}+t_n)/2$.
$G_{n\mp\frac12}$ live on the links immediately
before and after $t_n$, respectively, and $(\delta_\epsilon G)_n$ is site-centered at $t_n$.
Assuming that $G = 0$ near $t_\mathrm{i}$ and $t_\mathrm{f}$, a sufficiently small $\epsilon$ gives $G_{\frac12}=G_{N-\frac12}=0$, and therefore
\begin{align}
  \epsilon\sum_{n=1}^{N-1}(\delta_\epsilon G)_n
  =G_{N-\frac12}-G_{\frac12}=0.
  \label{eq:IG-telescope}
\end{align}
We then rewrite the path integral by an insertion of zero as
\begin{align}
	K_{N}
	&=\left(\frac{M}{2\pi i\epsilon}\right)^{N/2}
	\int\prod_{n=1}^{N-1}\dd q_n\,
	e^{iS_{G,N}[q]},
  \quad
  S_{G,N}[q]
  =S_{0,N}[q]
    +\epsilon\sum_{n=1}^{N-1}(\delta_\epsilon G)_n,
    \label{eq:path_integral_with0}
\end{align}
which has the form of a path integral constructed from the original Lagrangian $L = L_0 + \mathrm{d}G/\mathrm{d}t$.
By construction, the total time derivative term $\epsilon \sum_n (\delta_\epsilon G)_n$ has no effect in Eq.~\eqref{eq:path_integral_with0}.
The confusions in the literature arise from how to show this fact in explicit calculations.
In practical calculations, we expand the total derivative term in the exponent perturbatively as
\begin{equation}
  K_{N}
  =\left(\frac{M}{2\pi i\epsilon}\right)^{N/2}
    \int\prod_{n=1}^{N-1}\dd q_n\,
    e^{iS_{0,N}[q]}
    \sum_{r=0}^{\infty}\frac{i^r}{r!}
    \left[\sum_{n=1}^{N-1}\epsilon(\delta_\epsilon G)_n\right]^r.
  \label{eq:path-integral-perturbative-expansion}
\end{equation}
This discretized definition provides a well-defined prescription for the insertions of operators in the path integral, including the total time derivative term.

\subsection{%
  \texorpdfstring{$\Tstar$}{T-star}-product from path integral}
\label{sec:finite-time-sewing-tstar}

We now see that Eq.~\eqref{eq:path-integral-perturbative-expansion} automatically leads to the $\Tstar$-product, which becomes particularly important for the insertion of the total time derivative term in the correlators.

We first recall the case where the correlation function involves only $q$.
Consider the correlator of $q$ at discretized times $t_{n_1},\ldots,t_{n_r}$,
\begin{equation}
  \left\langle
    q_{n_1}\cdots q_{n_r}
  \right\rangle_{0,N}
  \equiv \mathcal{N} \, \int\prod_{n=1}^{N-1}\dd q_n\,q_{n_1}\cdots q_{n_r}\,e^{iS_{0,N}[q]}
  ,
  \label{eq:q-lattice-correlator}
\end{equation}
which is defined by the path integral average with respect to the weight $e^{iS_{0,N}}$.
We have introduced the normalization factor $\mathcal{N}$ to ensure that $\langle 1\rangle_{0,N}=1$.
The corresponding continuum limit is given by
\begin{equation}
  \lim_{N\to\infty}
    \left\langle
    q_{n_1}\cdots q_{n_r}
  \right\rangle_{0,N}
  =
  \left\langle
    \mathcal{T}\,\hat q(t_1)\cdots\hat q(t_r)
  \right\rangle_0,
  \qquad
  t_{n_a}\to t_a .
  \label{eq:q-lattice-to-T}
\end{equation}
The $\mathcal{T}$-ordering on the right-hand side is inherited from the ordered product of short-time evolution operators in the discretized path integral.
Notice that quantities in the bracket on the left-hand side are $c$-numbers, while those on the right-hand side are operators, which we made clear by denoting operators with hats.

The correlators involving $\dot{q}$ are obtained from the $q$-correlators by finite differences, with $\dot q$ given by $(q_{n_i+1}-q_{n_i})/{\epsilon}$, as
\begin{align}
  &
  \left\langle
    q_{n_1}\cdots \frac{q_{n_i+1}-q_{n_i}}{\epsilon} \cdots q_{n_r}
  \right\rangle_{0,N}
  =
  \frac{1}{\epsilon}
  \left[
    \left\langle
      q_{n_1}\cdots q_{n_i+1} \cdots q_{n_r}
    \right\rangle_{0,N}
    -
    \left\langle
      q_{n_1}\cdots q_{n_i} \cdots q_{n_r}
    \right\rangle_{0,N}
  \right].
\end{align}
The continuum limit of this correlator gives the $\Tstar$-product as
\begin{align}
  \lim_{N\to\infty}
  \left\langle
    q_{n_1}\cdots \frac{q_{n_i+1}-q_{n_i}}{\epsilon} \cdots q_{n_r}
  \right\rangle_{0,N}
  &=
  \partial_{t_i}
  \left\langle
    \mathcal{T} \, \hat q (t_{1}) \cdots \hat q (t_i) \cdots \hat q (t_{r})
  \right\rangle_{0}
    \nonumber \\
    &= \left\langle
    \mathcal{T}^\star\,\hat q(t_1)\cdots\dot{\hat q}(t_i)\cdots\hat q(t_r)
  \right\rangle_0,
  \label{eq:q-lattice-to-Tstar}
\end{align}
where the time derivative on the right-hand side is taken after the $\mathcal{T}$-ordering is performed.

Now we consider the correlators involving the total time derivative term $\dd G/\dd t$.
In the following, we take $G = q^2 \dot q$ as an illustrative example.
Extension to more general cases is straightforward.
Insertion of $G$ into the correlator gives
\begin{align}
  \left\langle
    q_{n_1}\cdots G_{n_i+\frac12}\cdots q_{n_r}
  \right\rangle_{0,N}
    &=
  \left\langle
    q_{n_1}\cdots \bar q_{n_i+\frac12}^2\frac{\Delta q_{n_i + \frac12}}{\epsilon}\cdots q_{n_r}
  \right\rangle_{0,N} \nonumber \\
  &=  \frac{1}{4\epsilon}
  \left\langle
    q_{n_1}\cdots
    (q_{n_i+1}+q_{n_i})^2
    (q_{n_i+1}-q_{n_i})
    \cdots q_{n_r}
  \right\rangle_{0,N} \nonumber \\
  &=
  \sum_{\sigma_1, \sigma_2 = \pm}
  \frac{1}{4\epsilon}
  \left\langle
    q_{n_1} \cdots
    q_{\sigma_1} q_{\sigma_2} (q_+-q_-)
    \cdots q_{n_r}
  \right\rangle_{0,N},
\end{align}
with $q_+ = q_{n_i + 1}$ and $q_- = q_{n_i}$.
Taking the continuum limit with $t_{n_i + 1/2} \to t$, we obtain
\begin{align}
  &\lim_{N\to\infty}
  \left\langle
    q_{n_1}\cdots G_{n_i+\frac12}\cdots q_{n_r}
  \right\rangle_{0,N}
  \nonumber \\
  &=
  \frac14 \lim_{\epsilon\to0} \, \lim_{t' \to t} \Big[ \partial_{t'} \left\langle \mathcal T\, \cdots \hat q(t_+)\hat q(t_+)\hat q(t') \cdots \right\rangle_0
  + \partial_{t'} \left\langle \mathcal T\, \cdots \hat q(t_+)\hat q(t_-)\hat q(t') \cdots \right\rangle_0 \nonumber \\
  &\qquad\qquad\quad ~ + \partial_{t'} \left\langle \mathcal T\, \cdots \hat q(t_-)\hat q(t_+)\hat q(t') \cdots \right\rangle_0 
  + \partial_{t'} \left\langle \mathcal T\, \cdots \hat q(t_-)\hat q(t_-)\hat q(t') \cdots \right\rangle_0 \Big],
\end{align}
with $t_\pm = t \pm \epsilon/2$.
We denote this limit as the $\Tstar$-product for the $\hat G$ operator, namely
\begin{equation}
  \left\langle
    \Tstar \hat q (t_1) \cdots \hat G(t) \cdots \hat q (t_r)
  \right\rangle_{0}
  \equiv
  \frac14 \lim_{\epsilon\to0} \sum_{\sigma_1,\sigma_2 = \pm}
  \left\langle
    \Tstar\, \hat q(t_1) \cdots \hat q(t_{\sigma_1})\hat q(t_{\sigma_2}) \dot{\hat q} (t) \cdots \hat q (t_r)
  \right\rangle_0.
  \label{eq:Tstar-G-definition}
\end{equation}
By definition, the insertion of the total time derivative in the time-discretized path integral is given by
\begin{equation}
  \left\langle
    q_{n_1}\cdots (\delta_\epsilon G)_{n_i}\cdots q_{n_r}
  \right\rangle_{0,N}
  =
  \frac{1}{\epsilon}
    \left\langle
      q_{n_1}\cdots \left( G_{n_i+\frac12} - G_{n_i-\frac12} \right) \cdots q_{n_r}
    \right\rangle_{0,N}.
\end{equation}
The continuum limit is expressed in terms of Eq.~\eqref{eq:Tstar-G-definition} as
\begin{equation}
  \left\langle
    \Tstar\, \hat q (t_1) \cdots \dot{\hat G}(t) \cdots \hat q (t_r)
  \right\rangle_{0}
  =
  \partial_t \left\langle
    \Tstar\, \hat q (t_1) \cdots \hat G(t) \cdots \hat q (t_r)
  \right\rangle_{0}.
\end{equation}
Recalling that the total time derivative is accompanied by the time integral as a continuum limit of the sum over $t_n$ in Eq.~\eqref{eq:path-integral-perturbative-expansion}, we obtain
\begin{align}
  \int \dd t\,
  \left\langle
    \Tstar\, \hat q (t_1) \cdots \dot{\hat G}(t) \cdots \hat q (t_r)
  \right\rangle_{0}
  =
  \int \dd t\, \partial_t \left\langle
    \Tstar\, \hat q (t_1) \cdots \hat G(t) \cdots \hat q (t_r)
  \right\rangle_{0}
  = 0.
  \label{eq:multi-Tstar-integrated-vanishing}
\end{align}
Here it is essential that the time derivative acts outside of the time ordering due to the $\mathcal{T}^\star$-product.
Therefore, the total derivative gives no contribution as long as the boundary term vanishes, which is guaranteed by our assumption.

The $\Tstar$-product is a well-known, textbook-level property of the path integral formulation in the in-out formalism.
See, for instance,~\cite{Kugo:1997gva,Bastianelli:2006rx,Itzykson:1980rh,Weinberg:1995mt,Srednicki:2007qs} and footnote~\ref{foot:Tstar} for notational differences among the textbooks.
Its extension to the in-in formalism naturally introduces the $\TCstar$-product along the closed time path, as we see now.

\section{Closed time path and \texorpdfstring{$\TCstar$}{TC-star}-product}

\label{sec:closed-time-path}

\begin{figure}
    \centering
    \includegraphics[width=0.8\textwidth]{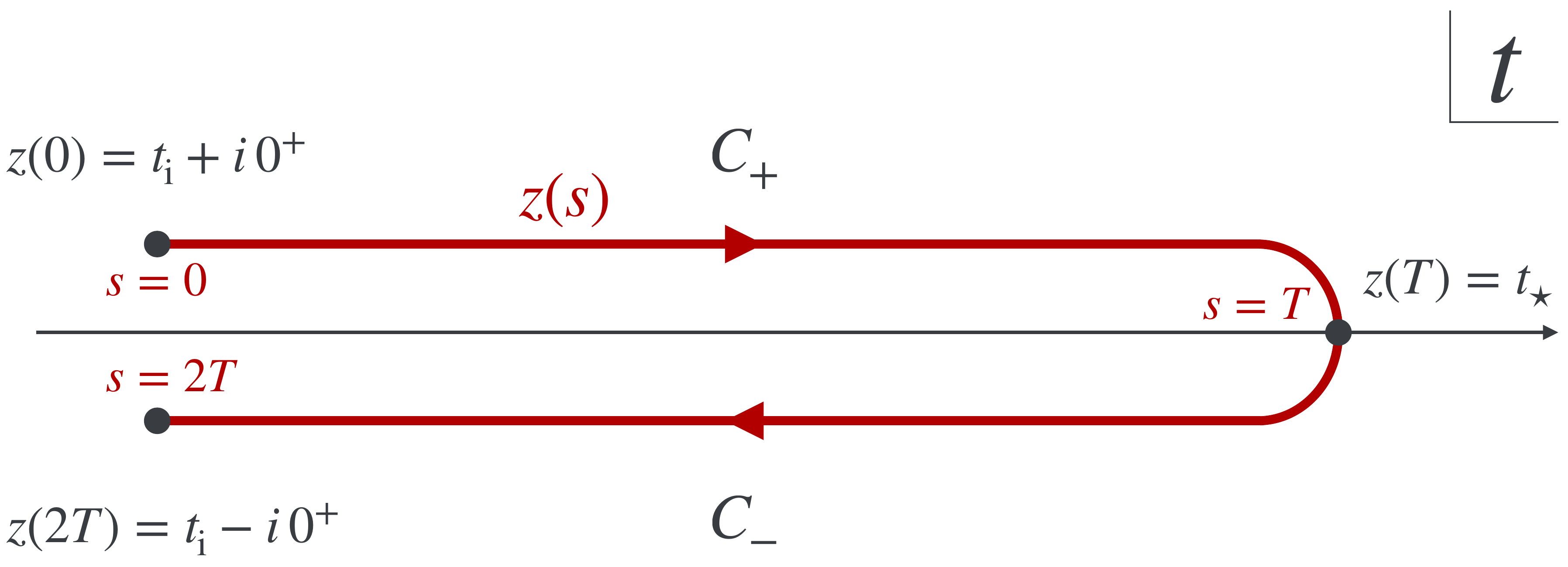}
    \caption{Closed time path, $C = C_+ + C_-$, in the complex-time plane.}
    \label{fig:ctp-contour}
\end{figure}

In this section, we apply the discrete construction in Sec.~\ref{sec:path-integral-tstar} to the closed time path denoted by $C$, which runs from $t_\mathrm{i}$ to $t_\star$ along the forward branch $C_+$ and back to $t_\mathrm{i}$ along the backward branch $C_-$, with $t_\star$ being the largest operator-insertion time as shown in Fig.~\ref{fig:ctp-contour}. 
We first write $\hat{U}_0(t_\star,t_\mathrm{i})\hat{\rho}_{\mathrm{i}}\hat{U}_0^\dagger(t_\star,t_\mathrm{i})$ as a discretized path integral and insert the total derivative as an exact zero.
We then take the continuum limit and derive the corresponding $\TCstar$-ordering.

\subsection{Total time derivative in the in-in path integral}
\label{sec:in-in-path-integral}

The in-in formalism computes expectation values of observables at a finite time from an initial state specified at $t_\mathrm{i}$. 
The expectation value of an observable $O$ evaluated at $t_\star$ is given by
\begin{equation}
\langle \hat O(t_\star)\rangle
=
\operatorname{Tr}\!\left[
\hat U_0(t_\star,t_\mathrm{i})\,
\hat \rho_{\mathrm{i}}\,
\hat U_0^\dagger(t_\star,t_\mathrm{i})\,
\hat O (t_\mathrm{i})
\right],
\end{equation}
where $\hat\rho_{\mathrm{i}}$ is the initial density matrix and, again, the system evolves with $\hat{H}_0$ as a result of Sec.~\ref{sec:canonical-quantization}.
We use the same time discretization as in Sec.~\ref{sec:path-integral-canonical} for $\hat U_0$ and $\hat U_0^\dagger$ with
\begin{equation}
t_n=t_\mathrm{i}+n\epsilon,
\qquad
\epsilon=\frac{t_\star-t_\mathrm{i}}{N},
\qquad
n=0,\ldots,N,
\end{equation}
and denote the configuration variables on the forward and backward branches by $q_n^+$ and $q_n^-$, respectively. 
In terms of the finite-$N$ kernel $K_{N}$, the expectation value of $\hat{q}(t_\star)$ is given by
\begin{align}
\langle q_N \rangle_{0,N}
&=
\int
\dd q_{0}^+\,
\dd q_{0}^-\,
\dd q_N \,
\rho_{\mathrm{i}}(q_{0}^+,q_{0}^-)\, q_N \,
K_{N}(q_N ,t_N;q_{0}^+,t_{0})\,
K_{N}^*(q_N,t_N; q_{0}^-,t_{0}).
\end{align}

Let us write this expression as a closed time path~\cite{Landsman:1986uw,Berges:2004yj,Haehl:2016pec,Liu:2018kfw,Haehl:2024pqu}.
We consider a smooth contour $C$ in the complex-time plane parametrized by a monotonically increasing $s$ (see Fig.~\ref{fig:ctp-contour}):
\begin{align}
    C =\{z(s)\mid 0\leq s\leq 2T\},
    \quad z'(s)\neq0, \quad \operatorname{Im} z'(s) \leq 0,
  \label{eq:ctp-parametrization}
\end{align}
where the parameter $s$ fixes contour ordering and
\begin{align}
    z(0)=t_\mathrm{i} + \ii 0^+, \quad z(2T)=t_\mathrm{i} - \ii 0^+,
    \quad z(T)=t_\star,
\end{align}
with $T\equiv t_\star-t_\mathrm{i}$.
We slightly extend the contour into the complex plane so that $z'(s)$ remains continuous and nonzero at the turning point while satisfying the sufficient condition for the convergence, $\operatorname{Im}z'(s)\leq0$.
At the end of the calculation, we flatten the contour to the real axis by taking $\operatorname{Im}z(s)\to0$ as in the $i\epsilon$ prescription.

We discretize the closed time path as
\begin{equation}
  s_n=n\epsilon,
  \qquad
  \epsilon=\frac{T}{N}>0,
  \qquad
  n=0,\ldots,2N,
\end{equation}
and define
\begin{equation}
  z_n\equiv z(s_n),
  \qquad
  q_n^C\equiv q(s_n).
\end{equation}
In the flattened limit, the branch variables satisfy $q_n^C=q_n^+$ for $n=0,\ldots,N$ and $q_n^C=q_{2N-n}^-$ for $n=N,\ldots,2N$, with $q_N^C=q_N^+=q_N^-=q_N$.
With the link variables
\begin{equation}
    \Delta q_{n+\frac12}^C\equiv q_{n+1}^C-q_n^C,
    \qquad
    \bar q_{n+\frac12}^C\equiv\frac{q_{n+1}^C+q_n^C}{2},
\end{equation}
the action along the closed time path is defined as
\begin{align}
    S_{0,N}[q^C] &= \sum_{n=0}^{2N-1} \epsilon_{n+ \frac12}^C \left[ \frac{M}{2} \left( \frac{\Delta q_{n+ \frac12}^C}{\epsilon_{n+ \frac12}^C} \right)^2 - V(\bar q_{n + \frac12}^C) \right]
    \xrightarrow{\operatorname{Im} z(s)\to0}
    S_{0,N}[q^+]-S_{0,N}[q^-],
\end{align}
where $\epsilon_{n+\frac12}^C \equiv z_{n+1}-z_n~(\neq0)$.
With this contour parametrization, the in-in path integral is expressed as
\begin{equation}
    \langle q_N^C \rangle_{0,N}
    =
    \mathcal{N} \int \prod_{n=0}^{2 N} \dd q_n^C \,
    \rho_{\mathrm{i}}(q_{0}^C,q_{2N}^C)\,
    q_N^C \,
    \exp \left(
        i
        S_{0,N}[q^C]
    \right),
\end{equation}
with
\begin{equation}
    \langle q_N \rangle_{0,N}
    =
    \lim_{\operatorname{Im} z(s)\to0}
    \langle q_N^C \rangle_{0,N}.
\end{equation}
This expression makes explicit the fact that the in-in path integral is a single path integral along the closed time path $C$, rather than two independent path integrals along the forward and backward branches.

The contour total derivative
$\int_C\dd z\,\dd G/\dd z=\int_0^{2T}\dd s\,\dd G/\dd s$ is represented as
\begin{align}
    G_{n+\frac12}^C &\equiv G\!\left(\bar q_{n+\frac12}^C, v_{n+\frac12}^C, z_{n+\frac12}\right),
    \quad n = 0, \cdots, 2N-1,
    \\
    (\delta_\epsilon G)_n^C &\equiv \frac{G_{n+\frac12}^C - G_{n-\frac12}^C}{\epsilon},
    \quad n=1,\cdots,2N-1,
\end{align}
where
\begin{equation}
    v_{n+\frac12}^C \equiv \frac{q_{n+1}^C - q_n^C}{\epsilon_{n+\frac12}^C}, \qquad
    z_{n+\frac12} \equiv z (s_{n+\frac12}), \qquad s_{n+\frac12} \equiv \frac{s_n+s_{n+1}}{2}.
\end{equation}
The sum satisfies
\begin{equation}
    \epsilon \sum_{n=1}^{2N-1} (\delta_\epsilon G)_n^C
    =
    \sum_{n=1}^{2N-1} (G_{n+\frac12}^C - G_{n-\frac12}^C)
    =
    G_{2N-\frac12}^C - G_{\frac12}^C =0 ,
\end{equation}
where a sufficiently small $\epsilon$ gives $G_{\frac12}^C=G_{2 N-\frac12}^C =0 $ under $G = 0$ near $t_\mathrm{i}$.
This again allows us to add the total time derivative to the in-in path integral action as
\begin{equation}
    S_{G,N} [q^C] = S_{0,N}[q^C] + \epsilon \sum_{n=1}^{2N-1} (\delta_\epsilon G)_n^C.
\end{equation}

\subsection{%
\texorpdfstring{$\TCstar$}{TC-star}-product}
\label{sec:tcstar-product}

In-in correlators are defined in a similar way to those in the in-out formalism.
The finite-$N$ correlator is given by
\begin{equation}
    \langle q_{n_1}^C \cdots q_{n_r}^C \rangle_{0,N}
    =
    \mathcal{N} \int \prod_{n=0}^{2 N} \dd q_n^C \,
    \rho_{\mathrm{i}}(q_{0}^C,q_{2N}^C)\,
    q_{n_1}^C \cdots q_{n_r}^C \,
    \exp \left(
       iS_{0,N}[q^C]
    \right),
\end{equation}
and its continuum limit gives the contour-ordered correlation function along the closed time path $C$ as
\begin{equation}
    \lim_{N\to\infty}
    \langle q_{n_1}^C \cdots q_{n_r}^C \rangle_{0,N}
    =
    \langle \TC \, \hat q(s_1) \cdots \hat q(s_r) \rangle_0.
\end{equation}
Here $s_a$ is the continuum limit of $s_{n_a}$ and $z_a=z(s_a)$ is its time coordinate.
The operator $\TC$ orders insertions by $s_a$, distinguishing contour points even when their time coordinates agree.

Correlators involving time derivatives are obtained in the same way as in Sec.~\ref{sec:finite-time-sewing-tstar}.
The discretized expression of the derivative with respect to $z$ is
\begin{equation}
  \frac{q_{n+1}^C-q_n^C}{z_{n+1}-z_n}
  =
  \frac{q_{n+1}^C-q_n^C}{\epsilon_{n+\frac12}^C},
\end{equation}
and its correlator is
\begin{align}
  &
  \left\langle
    q_{n_1}^C\cdots
    \frac{q_{n_i+1}^C-q_{n_i}^C}{\epsilon_{n_i+\frac12}^C}
    \cdots q_{n_r}^C
  \right\rangle_{0,N}
  =
  \frac{1}{\epsilon_{n_i+\frac12}^C}
  \left[
    \left\langle
      q_{n_1}^C\cdots q_{n_i+1}^C\cdots q_{n_r}^C
    \right\rangle_{0,N}
    -
    \left\langle
      q_{n_1}^C\cdots q_{n_i}^C\cdots q_{n_r}^C
    \right\rangle_{0,N}
  \right].
\end{align}
Its continuum limit defines the $\TCstar$-product as
\begin{align}
  \left\langle
    \TCstar\,
    \hat q(s_1)\cdots
    \dot{\hat q}(s_i)
    \cdots\hat q(s_r)
  \right\rangle_0
  &=
  \partial_{z_i}
  \left\langle
    \TC\,
    \hat q(s_1)\cdots
    \hat q(s_i)
    \cdots\hat q(s_r)
  \right\rangle_0,
  \label{eq:q-contour-to-TCstar}
\end{align}
where the derivative acts outside of the contour-ordering.

In the same way as in Sec.~\ref{sec:finite-time-sewing-tstar}, we illustrate the insertion of $G$ with $G=q^2\dot q$.
The time-discretized correlator is given by
\begin{equation}
    \left\langle
        q_{n_1}^C \cdots G_{n_i +\frac12}^C \cdots q_{n_r}^C
    \right\rangle_{0,N}
    =
    \frac{1}{4 \epsilon_{n_i+\frac12}^C}
    \left\langle
        q_{n_1}^C \cdots \left( q_{n_i+1}^C + q_{n_i}^C \right)^2 \left( q_{n_i+1}^C - q_{n_i}^C \right) \cdots q_{n_r}^C
    \right\rangle_{0,N}.
\end{equation}
Taking the continuum limit with $s_{n_i + 1/2} \to s$, we obtain the $\mathcal T_C^\star$-ordered correlator with the insertion of $G$ in analogy with Eq.~\eqref{eq:Tstar-G-definition} as
\begin{equation}
    \left\langle
        \TCstar\, \hat q (s_1) \cdots \hat G (s) \cdots \hat q (s_r)
    \right\rangle_{0}
    =
    \frac{1}{4} \lim_{\epsilon \to 0}
    \sum_{\sigma_1,\sigma_2 = \pm}
  \left\langle
    \TCstar\, \hat q(s_1) \cdots \hat q(s_{\sigma_1})\hat q(s_{\sigma_2}) \dot{\hat q} (s) \cdots \hat q (s_r)
  \right\rangle_0,
\end{equation}
with $s_{\pm} = s \pm \epsilon/2$.

Finally, we show that the total time derivative in the in-in path integral does not contribute after the time integration as long as $G = 0$ near $t_\mathrm{i}$.
By definition, the insertion of $(\delta_\epsilon G)_n^C$ in the discretized path integral is given by
\begin{equation}
    \left\langle
        q_{n_1}^C \cdots (\delta_\epsilon G)_{n_i}^C \cdots q_{n_r}^C
    \right\rangle_{0,N}
    =
    \frac{1}{\epsilon}
        \left\langle
            q_{n_1}^C \cdots \left( G_{n_i+\frac12}^C - G_{n_i-\frac12}^C \right) \cdots q_{n_r}^C
        \right\rangle_{0,N}.
\end{equation}
The corresponding continuum limit gives the $\TCstar$-product as
\begin{equation}
    \left\langle
        \TCstar\, \hat q (s_1) \cdots \partial_s \hat G (s) \cdots \hat q (s_r)
    \right\rangle_{0}
    =
    \partial_s 
    \left\langle
        \TCstar\, \hat q (s_1) \cdots \hat G (s) \cdots \hat q (s_r)
    \right\rangle_{0}.
\end{equation}
Since the insertion of $\dd G/\dd z$ is accompanied by the time integral, we obtain
\begin{align}
    \int_C \dd z \,
    \left\langle
        \TCstar\, \hat q (s_1) \cdots \dot{\hat G} (s) \cdots \hat q (s_r)
    \right\rangle_{0}
    &=
    \int_0^{2T} \dd s \,
    \left\langle
        \TCstar\, \hat q (s_1) \cdots \partial_s \hat G (s) \cdots \hat q (s_r)
    \right\rangle_{0} \nonumber \\
    &=
    \int_0^{2T} \dd s \,\partial_s
    \left\langle
        \TCstar\, \hat q (s_1) \cdots  \hat G (s) \cdots \hat q (s_r)
    \right\rangle_{0}
    = 0.
\end{align}
This proof in the in-in path integral is equivalent to the one in Sec.~\ref{sec:path-integral-tstar} for the in-out path integral, once we notice that the correct operator ordering is the $\TCstar$-ordering.

The notion of the $\TCstar$-product is crucial here.
Aside from the turning point, the standard $\Tstar$-/$\bar{\mathcal{T}}^\star$-products in the upper and lower branches are enough.
However, for the operator insertion exactly at the turning point, the $\TCstar$-product is required to correctly account for the contact terms (see Sec.~\ref{sec:examples} for details).
Once we recall that the in-in contour is nothing but a single contour $C$ in the complex-time plane, the $\TCstar$-product is understood as a natural generalization of the $\Tstar$-product, and then the vanishing of the total time derivative term in the in-in path integral is as straightforward as in the in-out path integral.

\section{Explicit calculations with \texorpdfstring{$\TCstar$}{TC-star}-product}
\label{sec:examples}

We saw in Sec.~\ref{sec:closed-time-path} that the path integral formulation leads to the $\TCstar$-product in the in-in formalism.
In the $\TCstar$-product, time derivatives act outside of contour orderings, and therefore a total time derivative reduces to the contributions at the endpoints of the path integral contour.
The in-in contour $C$ is defined as one smooth contour with the endpoints at $\tau = -\infty \pm i0^+$ and the evaluation point $\tau = \tau_\star$ is never an endpoint, in the same way that $\tau = \tau_\star$ is not a special point in the in-out formalism.
Thus, the total derivative term does not give rise to any contribution localized at $\tau = \tau_\star$.

In this section, to see how our discussion corrects previous literature that claims the existence of the total time derivative contributions at $\tau = \tau_\star$,
we take the correlation functions considered in~\cite{Braglia:2024zsl} as an example.
We will see that the $\TCstar$-product introduces additional delta function-type contributions to correlators that were previously omitted.
By including these contributions, we restore the cancellation of the total time derivative terms.

\subsection{Three-point function}
\label{subsec:3pt_explicit}

Let us consider a massless scalar field $\phi$ with the following action
\begin{align}
	S = \int \dd\tau \dd^{d-1}x\,
    \left[\frac{f^2(\tau)}{2} \left( (\partial_\tau \phi)^2 - (\partial_i \phi)^2 \right) +
    \frac{\dd}{\dd\tau}\left(f^2(\tau)\phi^2 \partial_\tau \phi\right)
    \right],
    \label{eq:action_explicit}
\end{align}
where $f(\tau)$ is an arbitrary function.\footnote{
With only one time derivative, the $\Tstar$-product reduces to the naive $\mathcal{T}$-product because  $G_>(k;\tau_1,\tau_1) = G_<(k;\tau_1,\tau_1)$ ensures that the contributions from differentiating the theta functions cancel so that $\partial_{\tau_1} G_F(k;\tau_1,\tau_2) = \theta(\tau_1-\tau_2) \partial_{\tau_1}G_>(k;\tau_1,\tau_2) + \theta(\tau_2-\tau_1) \partial_{\tau_1}G_<(k;\tau_1,\tau_2)$.
The same is true for the $\TCstar$-product, and therefore we focus on the total time derivative term that contains $\partial_\tau \phi$, where the $\Tstar/\TCstar$-product plays a nontrivial role even for the three-point function.
}
We take the coefficient inside the total time derivative to be $f^2(\tau)$, but we can choose it arbitrarily without affecting the discussion below.
We quantize $\phi$ as
\begin{align}
	\hat{\phi}(\tau,\vec{x}) = \frac{1}{f}\int\frac{\dd^{d-1}k}{(2\pi)^{d-1}}\,e^{i\vec{k}\cdot \vec{x}}
	\left[\phi_{\vec{k}}(\tau) \hat{a}_{\vec{k}} + \phi^*_{-\vec{k}}(\tau)\hat{a}^\dagger_{-\vec{k}}\right],
\end{align}
where the mode function satisfies the equation of motion
\begin{align}
	\left[\frac{\dd^2}{\dd\tau^2} + k^2 - \frac{f''}{f}\right]\phi_{\vec{k}}(\tau) = 0,
\end{align}
and the Wronskian condition
\begin{align}
    \phi_k(\tau)\partial_\tau \phi_k^*(\tau) - \phi_k^*(\tau) \partial_\tau \phi_k(\tau) = i.
    \label{eq:Wronskian}
\end{align}
We calculate the three-point correlators evaluated at $\tau = \tau_\star$ in this subsection.

\subsubsection*{In-out formalism}

As a warm-up, let us calculate the three-point correlators evaluated at $\tau = \tau_\star$ in the in-out formalism, where there is no risk of confusion about the absence of total time derivative contributions.
We now demonstrate that, to achieve this well-known result, we need to properly include delta function-type contributions arising from the $\mathcal{T}^\star$-product.

The Feynman propagator in the in-out formalism is given by
\begin{align}
    G_F(k; \tau_1,\tau_2) = \theta(\tau_1-\tau_2) G_>(k; \tau_1,\tau_2) + \theta(\tau_2 - \tau_1) G_<(k; \tau_1,\tau_2),
\end{align}
where
\begin{align}
    G_>(k; \tau_1, \tau_2) =  \frac{\phi_{k}(\tau_1)\phi_k^*(\tau_2)}{f(\tau_1)f(\tau_2)},
    \quad
    G_<(k; \tau_1, \tau_2) =  \frac{\phi_{k}(\tau_2)\phi_k^*(\tau_1)}{f(\tau_1)f(\tau_2)}.
\end{align}
The three-point function at the leading order in perturbations, arising from the total time derivative term, is expressed as
\begin{align}
     &\langle \hat{\phi}_{k_1}(\tau_\star) \hat{\phi}_{k_2}(\tau_\star) \hat{\phi}_{k_3}(\tau_\star) \rangle'
     \nonumber \\
     &= 2i \int_{-\infty}^{\infty} \dd\tau\, \partial_{\tau} \Big[f^2(\tau) G_F(k_2;\tau_\star,\tau)G_F(k_3;\tau_\star,\tau)\partial_{\tau} G_F(k_1;\tau_\star,\tau)
     + (\mathrm{perms.}) \Big],
     \label{eq:3pt_in-out}
\end{align}
where the prime on the correlator indicates that the delta function of the total momentum conservation is stripped out, 
``(perms.)'' indicates the permutations $k_1 \leftrightarrow k_2, k_3$, 
and the time derivative acts outside of the theta functions due to the $\mathcal{T}^\star$-product.
This is a total time derivative contribution and hence vanishes as long as the contributions are suppressed at $\tau \to \pm \infty$ by, \textit{e.g.}, the $i\epsilon$-prescription.
In particular, there is no nontrivial contribution at the operator insertion point, $\tau = \tau_\star$.

As an illustration, let us take the time derivatives explicitly.
Taking care of the derivatives acting on the theta functions, we obtain
\begin{align}
    &\partial_{\tau} \Big[f^2(\tau) G_F(k_2;\tau_\star,\tau)G_F(k_3;\tau_\star,\tau)\partial_{\tau} G_F(k_1;\tau_\star,\tau) \Big]
    \nonumber \\
    &= \sum_{\lambda=\pm}\theta(\lambda(\tau_\star-\tau)) \partial_\tau\Big[f^2(\tau) G_{\lambda}(k_2;\tau)G_\lambda(k_3;\tau) \partial_\tau G_{\lambda}(k_1;\tau)\Big]
    - i\delta(\tau-\tau_\star) P(k_2) P(k_3),
\end{align}
where we defined
\begin{align}
    G_+(k;\tau) = G_>(k;\tau_\star,\tau),
    \quad
    G_-(k;\tau) = G_<(k;\tau_\star,\tau),
    \quad
    G_>(k;\tau_\star,\tau_\star) = G_<(k;\tau_\star,\tau_\star) = P(k),
\end{align}
and used the Wronskian condition
\begin{align}
    \Big[\partial_\tau G_>(k;\tau_\star,\tau) - \partial_\tau G_<(k;\tau_\star,\tau)\Big]_{\tau \to \tau_\star} = \frac{i}{f^2(\tau_\star)}.
\end{align}
Substituting this expression into Eq.~\eqref{eq:3pt_in-out}, we obtain
\begin{align}
    &\langle \hat{\phi}_{k_1}(\tau_\star) \hat{\phi}_{k_2}(\tau_\star) \hat{\phi}_{k_3}(\tau_\star) \rangle'
    \nonumber \\
    &= 2i\sum_{\lambda=\pm}\lambda \int_{\lambda} \dd\tau\, \partial_\tau \Big[f^2(\tau) G_\lambda(k_2;\tau) G_\lambda(k_3;\tau) \partial_\tau G_\lambda(k_1;\tau)\Big]
    + 2P(k_2)P(k_3) + (\mathrm{perms.}),
    \label{eq:3pt_in-out_split}
\end{align}
where we use the notation
\begin{align}
    \int_+\dd\tau \equiv \int_{-\infty}^{\tau_\star}\dd\tau,
    \quad
    \int_-\dd\tau \equiv \int_{+\infty}^{\tau_\star} \dd\tau,
\end{align}
to make the relation between Eq.~\eqref{eq:3pt_in-out_split} and the in-in formalism transparent.
If we split the in-out path integral contour $C$ into $C = C_+ + C_-$ with $C_+$ corresponding to $\tau \in [-\infty,\tau_\star]$ and $C_-$ to $\tau \in [\tau_\star,\infty]$, the first term in Eq.~\eqref{eq:3pt_in-out_split} arises from the naive $\mathcal{T}$-product after the split, while the second term in Eq.~\eqref{eq:3pt_in-out_split} arises from the delta function associated with the $\mathcal{T}^\star$-product.
To obtain the correct result, we need to include both contributions.
Indeed, the former contribution is evaluated as 
\begin{align}
    &2i\sum_{\lambda=\pm}\lambda \int_{\lambda} \dd\tau\, \partial_\tau \Big[f^2(\tau) G_\lambda(k_2;\tau) G_\lambda(k_3;\tau) \partial_\tau G_\lambda(k_1;\tau)\Big]
    \nonumber \\
    &= 2i f^2(\tau_\star)P(k_2)P(k_3) \sum_{\lambda=\pm}\lambda \Big[\partial_\tau G_\lambda(k_1;\tau)\Big]_{\tau\to\tau_\star}
    = -2 P(k_2) P(k_3),
\end{align}
which does not vanish.
However, after adding the second contribution in Eq.~\eqref{eq:3pt_in-out_split}, we obtain
\begin{align}
    \langle \hat{\phi}_{k_1}(\tau_\star) \hat{\phi}_{k_2}(\tau_\star) \hat{\phi}_{k_3}(\tau_\star) \rangle = 0.
\end{align}
Our calculation here is just to clarify that the $\Tstar$-product introduces additional contact terms compared to the naive $\mathcal{T}$-product; there is no need to take the derivatives explicitly in the actual calculations since the whole contribution trivially vanishes due to the original expression~\eqref{eq:3pt_in-out} that follows from the $\Tstar$-product. 

\subsubsection*{In-in formalism}

Let us now move to the in-in formalism.
Since the path integral leads to the $\TCstar$-product, we define the generalized Feynman propagator along the in-in contour $C$, given by
\begin{align}
    G_C(k; \tau_1,\tau_2) = \theta_C(\tau_1-\tau_2) G_>(k; \tau_1,\tau_2) + \theta_C(\tau_2 - \tau_1) G_<(k; \tau_1,\tau_2),
    \label{eq:contour_Feynman}
\end{align}
where $\theta_C$ is defined along the contour $C$, \emph{i.e.}, $\theta_C(\tau_1-\tau_2) = 1\, (= 0) $ if $\tau_1$ ($\tau_2$) comes later than $\tau_2$ ($\tau_1$) along the contour $C$.
In other words, $\theta_C$ corresponds to the ordinary theta function with respect to the parameter $s$ introduced in Sec.~\ref{sec:closed-time-path}.
Its derivative is given by
\begin{align}
    \partial_{\tau_1}\theta_C(\tau_1-\tau_2) = \delta_C(\tau_1-\tau_2),
    \label{eq:contour_theta_delta}
\end{align}
where $\delta_C(\tau_1-\tau_2)$ is the contour delta function having support only when $\tau_1$ and $\tau_2$ are at the same point on the contour $C$ (in this sense, the points on the forward and backward branches are distinct). It satisfies
\begin{align}
    \int_C\dd\tau\, \delta_C(\tau-\tau_1) f(\tau) = f(\tau_1),
\end{align}
where the time integral is performed along the in-in contour $C$.
Without time derivatives acting on it, the contour Feynman propagator~\eqref{eq:contour_Feynman} reduces to the standard four types of in-in propagators after decomposing $C = C_+ + C_-$ depending on whether $\tau_1, \tau_2$ lie on $C_\pm$.

The three-point function in the in-in formalism at the leading order in perturbations is written as 
\begin{align}
    &\langle \hat{\phi}_{k_1}(\tau_\star) \hat{\phi}_{k_2}(\tau_\star) \hat{\phi}_{k_3}(\tau_\star) \rangle'
    \nonumber \\
    &= 2i \int_C \dd\tau\, \partial_{\tau} \Big[f^2(\tau) G_C(k_2;\tau_\star,\tau)G_C(k_3;\tau_\star,\tau)\partial_{\tau} G_C(k_1;\tau_\star,\tau)
    + (\mathrm{perms.}) \Big].
\end{align}
Since this is a total derivative along the contour $C$, it reduces to the contributions at $\tau \to -\infty \pm i0^+$, which are suppressed by the $i\epsilon$-prescription, and not at $\tau \to \tau_\star$.

To see that this expression contains additional terms overlooked in the previous studies, let us explicitly take the time derivatives.
Including the derivatives acting on $\theta_C$, we obtain
\begin{align}
    &\partial_{\tau} \Big[f^2(\tau) G_C(k_2;\tau_\star,\tau)G_C(k_3;\tau_\star,\tau)\partial_{\tau} G_C(k_1;\tau_\star,\tau)  \Big]
    \nonumber \\
    &= 
    \sum_{\lambda=\pm} \theta_C(\lambda(\tau_\star-\tau))\partial_\tau\Big[f^2(\tau) G_{\lambda}(k_2;\tau)G_{\lambda}(k_3;\tau)\partial_\tau G_\lambda(k_1;\tau)\Big]
    - i\delta_C(\tau-\tau_\star) P(k_2) P(k_3),
\end{align}
where we used the Wronskian condition.
If we decompose $C = C_+ + C_-$, $\theta_C(\tau_\star-\tau)$ projects the time integral onto $C_+$, while $\theta_C(\tau-\tau_\star)$ projects onto $C_-$. Substituting this expression, we obtain
\begin{align}
    &\big\langle \hat{\phi}_{k_1}(\tau_\star) \hat{\phi}_{k_2}(\tau_\star) \hat{\phi}_{k_3}(\tau_\star) \big\rangle'
    \nonumber \\
    &= 2i \sum_{\lambda=\pm} \lambda\int_{-\infty}^{\tau_\star} \dd\tau\,
    \partial_{\tau} \Big[f^2(\tau) G_\lambda(k_2;\tau)G_{\lambda}(k_3;\tau)\partial_{\tau} G_\lambda(k_1;\tau)\Big] + 2 P(k_2) P(k_3) + (\mathrm{perms.}).
    \label{eq:3pt_in-in_split}
\end{align}
Note that this is in one-to-one correspondence with Eq.~\eqref{eq:3pt_in-out_split} in the in-out formalism.
The first term in Eq.~\eqref{eq:3pt_in-in_split} corresponds to the naive contribution from $C_+$ and $C_-$, while the second term arises from the delta function as a result of the $\TCstar$-product.\footnote{
    Ref.~\cite{Braglia:2024zsl} included only the first term in Eq.~\eqref{eq:3pt_in-in_split}.
    The same comment applies to the four-point examples we calculate below.
}
Indeed, the first term results in
\begin{align}
    2i \sum_{\lambda=\pm} \lambda\int_{-\infty}^{\tau_\star} \dd\tau\,
    \partial_{\tau} \Big[f^2(\tau) G_\lambda(k_2;\tau)G_{\lambda}(k_3;\tau)\partial_{\tau} G_\lambda(k_1;\tau)\Big]
    &= -2 P(k_2) P(k_3),
\end{align}
which is nonzero.
However, the second term in Eq.~\eqref{eq:3pt_in-in_split} provides an additional contribution so that the sum vanishes
\begin{align}
    &\big\langle \hat{\phi}_{k_1}(\tau_\star) \hat{\phi}_{k_2}(\tau_\star) \hat{\phi}_{k_3}(\tau_\star) \big\rangle = 0.
\end{align}
Thus, our calculation demonstrates that, after including the delta function arising from the $\TCstar$-product, the total time derivative does not affect the correlator.

As evident, the calculations here have one-to-one correspondence with those in the in-out formalism. Therefore, omitting the delta functions arising from the $\TCstar$-product is equivalent to omitting the corresponding delta functions from the $\mathcal{T}^\star$-product in the in-out formalism, contradicting the fact that the total derivative terms do not affect perturbations in the in-out formalism.

\subsubsection*{Equation-of-motion term}

Although our main focus in this paper is the total time derivative terms, here we comment on the terms that contain the equation-of-motion operator.
This operator contains second time derivatives, and therefore we need the same care about the additional delta functions arising from the $\TCstar$-product.

To be concrete, let us take the action as
\begin{align}
    S = \int \dd\tau \dd^{d-1}x\,
    \left[\frac{f^2(\tau)}{2} \left( (\partial_\tau \phi)^2 - (\partial_i \phi)^2 \right) +
    \frac{\dd}{\dd\tau}\left(f^2(\tau)\phi^2 \partial_\tau \phi\right)
    - f^2(\tau)\left(\mathcal{D}\phi\right) \phi^2
    \right],
    \label{eq:action_with_eom}
\end{align}
where the equation-of-motion operator $\mathcal{D}$ is defined as
\begin{align}
    \mathcal{D} =\frac{1}{f^2(\tau)}\frac{\dd}{\dd\tau}\left(f^2(\tau)\frac{\dd}{\dd\tau}\right) - \partial_i^2.
\end{align}
We can show by using Eqs.~\eqref{eq:Wronskian} and~\eqref{eq:contour_theta_delta} that the contour Feynman propagator satisfies
\begin{align}
    \mathcal{D}_{1} G_C(k;\tau_1,\tau_2) = -\frac{i}{f^2(\tau_1)}\delta_C(\tau_1-\tau_2),
    \label{eq:FeynmanC_eom}
\end{align}
where the subscript ``$1$'' indicates that it acts on $\tau_1$ and $-\partial_i^2$ is replaced by $k^2$ inside $\mathcal{D}_1$.
The three-point function originating from the equation-of-motion operator is given by
\begin{align}
    &\left.\langle \hat{\phi}_{k_1}(\tau_\star) \hat{\phi}_{k_2}(\tau_\star) \hat{\phi}_{k_3}(\tau_\star) \rangle'\right\vert_{\mathrm{EoM}}
    \nonumber \\
    &= -2i \int_C \dd\tau\, \Big[f^2(\tau) G_C(k_2;\tau_\star,\tau)G_C(k_3;\tau_\star,\tau)\mathcal{D} G_C(k_1;\tau_\star,\tau)
    + (\mathrm{perms.}) \Big],
\end{align}
where we note again that the equation-of-motion operator acts outside the contour theta functions as a result of the $\TCstar$-product.
Contrary to the claim in the previous study, this contribution does not vanish, and with Eq.~\eqref{eq:FeynmanC_eom} we obtain
\begin{align}
     \left.\langle \hat{\phi}_{k_1}(\tau_\star) \hat{\phi}_{k_2}(\tau_\star) \hat{\phi}_{k_3}(\tau_\star) \rangle'\right\vert_{\mathrm{EoM}}
     &= -2\Big[P(k_2)P(k_3) + P(k_3)P(k_1) + P(k_1)P(k_2)\Big],
\end{align}
which reproduces the final bispectrum in the previous literature~\cite{Braglia:2024zsl}.

This is not a coincidence.
We first explain why ordinary contour ordering gives the correct bispectrum when the two terms in Eq.~\eqref{eq:action_with_eom} are combined before evaluating the contractions.
The agreement on the final result can be traced back to Eq.~\eqref{eq:action_with_eom} in which the second time derivatives cancel out between the total time derivative term and the equation-of-motion term.
Since $G_>(k;\tau,\tau) = G_<(k;\tau,\tau)$, we see that
\begin{align}
    \partial_{\tau_1} G_C(k;\tau_1,\tau_2)
    = \theta_C(\tau_1-\tau_2)\partial_{\tau_1} G_>(k;\tau_1,\tau_2)
    + \theta_C(\tau_2-\tau_1) \partial_{\tau_1} G_<(k;\tau_1,\tau_2),
\end{align}
and therefore the $\TCstar$-product and the naive $\mathcal{T}_C$-product coincide for these external propagators as long as the action contains only first time derivatives.
On the other hand, when the two terms are evaluated separately, mistakes of omitting the delta functions arising from the $\TCstar$-product in both the total time derivative term and the equation-of-motion term therefore compensate for each other, resulting in the correct final result after the summation.
Nevertheless, the correct interpretation of the result is that it originates solely from the equation-of-motion term, and never from the total time derivative term.

\subsection{Four-point function}
\label{subsec:4pt_explicit}

In this subsection, we confirm the cancellation of the total time derivatives once again by calculating four-point correlators.
Since the calculations in the in-out and in-in formalisms are almost identical, we focus on the in-in formalism in the following.

We consider the contributions solely coming from the total time derivatives in Eq.~\eqref{eq:action_explicit} at the tree level.
There are three types of such (all vanishing) contributions, depending on whether the time derivative inside the total time derivative acts on the external or internal propagators.
We focus on those with at least one derivative acting on the external propagators, as the delta functions arising from the $\TCstar$-product play nontrivial roles in such cases.

\subsubsection*{Two derivatives on external propagators}

We first consider the contributions where both derivatives act on the external propagators.
\begin{align}
    \left.\big\langle \hat{\phi}_{k_1}(\tau_\star) \hat{\phi}_{k_2}(\tau_\star) \hat{\phi}_{k_3}(\tau_\star) \hat{\phi}_{k_4}(\tau_\star)\big\rangle'\right\vert_{2e}
    &= I_{2}(k_1,k_2;k_3,k_4) + (\mathrm{perms.}),
\end{align}
where the ``(perms.)'' indicates the remaining inequivalent permutations of the external momenta, and
the subscript ``$2e$'' indicates the contributions where both derivatives act on the external propagators.
Here we define $I_{2}$ as
\begin{align}
	&I_{2}(k_1,k_2;k_3,k_4) = -4\int_C\dd\tau_L \int_C\dd\tau_R\, \partial_{\tau_L}\partial_{\tau_R}\Big[f^2(\tau_L) f^2(\tau_R) 
	\nonumber \\
	&\times G_C(k_1;\tau_\star, \tau_L)  \partial_{\tau_L}G_C(k_2;\tau_\star, \tau_L) G_C(k_3;\tau_\star,\tau_R) \partial_{\tau_R}G_C(k_4;\tau_\star,\tau_R)
	  G_C(k_s;\tau_L,\tau_R)
	\Big],
    \label{eq:I2}
\end{align}
where $k_s = \vert \vec{k}_1 + \vec{k}_2\vert$.
Again, the derivative acts outside of the contour ordering theta functions, and it reduces to the boundary contributions at $\tau_{L,R} \to -\infty \pm i0^+$ which vanish trivially.

As before, let us take the time derivatives explicitly.
By a straightforward calculation, we obtain
\begin{align}
    I_2(k_1,&k_2;k_3,k_4)
    = -4\sum_{\lambda_L,\lambda_R=\pm}\lambda_L \lambda_R \int_{-\infty}^{\tau_\star} \dd\tau_L \int_{-\infty}^{\tau_\star} \dd\tau_R\, \partial_{\tau_L}\partial_{\tau_R}
	\nonumber \\
	&\times\Big[f_L^2f_R^2 G_{\lambda_L}(k_1;\tau_L)  \partial_{\tau_L}G_{\lambda_L}(k_2;\tau_L) G_{\lambda_R}(k_3;\tau_R) \partial_{\tau_R}G_{\lambda_R}(k_4;\tau_R)
	G_{\lambda_L \lambda_R}(k_s;\tau_L,\tau_R)
	\Big]
    \nonumber \\
    &+4iP(k_1) \sum_{\lambda_R = \pm} \lambda_R \int_{-\infty}^{\tau_\star} \dd\tau_R\,\partial_{\tau_R}\Big[f_R^2 G_{\lambda_R}(k_3;\tau_R) \partial_{\tau_R}G_{\lambda_R}(k_4;\tau_R) G_{\lambda_R}(k_s;\tau_R)\Big]
    \nonumber \\
    &+4iP(k_3) \sum_{\lambda_L = \pm} \lambda_L \int_{-\infty}^{\tau_\star} \dd\tau_L\, \partial_{\tau_L}\Big[f_L^2 G_{\lambda_L}(k_1;\tau_L) \partial_{\tau_L}G_{\lambda_L}(k_2;\tau_L) G_{\lambda_L}(k_s;\tau_L)\Big]
    \nonumber \\
    &+ 4P(k_1) P(k_3) P(k_s),
    \label{eq:I2_explicit}
\end{align}
where $f_{L/R} = f(\tau_{L/R})$ and
\begin{align}
    &G_{++}(k;\tau_1,\tau_2) = G_F(k;\tau_1,\tau_2),
    \quad
    G_{+-}(k;\tau_1,\tau_2) = G_<(k;\tau_1,\tau_2),
    \\
    &G_{-+}(k;\tau_1,\tau_2) = G_>(k;\tau_1,\tau_2),
    \quad
    G_{--}(k;\tau_1,\tau_2) = G_{\bar{F}}(k;\tau_1,\tau_2),
\end{align}
with $G_{\bar{F}}$ being the anti-Feynman propagator defined by the anti-time ordering $\bar{\mathcal{T}}$.
Again, the term in the first two lines in Eq.~\eqref{eq:I2_explicit} is the naive contribution from $C_+$ and $C_-$, while the remaining terms arise from the delta functions associated with the $\TCstar$-product.
The first term is evaluated as
\begin{align}
    &\sum_{\lambda_L,\lambda_R=\pm}\lambda_L \lambda_R \int_{-\infty}^{\tau_\star} \dd\tau_L \int_{-\infty}^{\tau_\star} \dd\tau_R\, \partial_{\tau_L}\partial_{\tau_R}
	\nonumber \\
    &\times\Big[f_L^2f_R^2 G_{\lambda_L}(k_1;\tau_L)  \partial_{\tau_L}G_{\lambda_L}(k_2;\tau_L) G_{\lambda_R}(k_3;\tau_R) \partial_{\tau_R}G_{\lambda_R}(k_4;\tau_R)
	G_{\lambda_L \lambda_R}(k_s;\tau_L,\tau_R)
	\Big]
    \nonumber \\
    &= -P(k_1)P(k_3) P(k_s),
\end{align}
where we used the Wronskian condition and $[\theta(\tau_L-\tau_R) + \theta(\tau_R-\tau_L)]_{\tau_L,\tau_R\to\tau_\star} = 1$.
This is again nonvanishing. However, the second and third terms are evaluated as
\begin{align}
    \sum_{\lambda_R = \pm} \lambda_R \int_{-\infty}^{\tau_\star} \dd\tau_R\,\partial_{\tau_R}\Big[f_R^2 G_{\lambda_R}(k_3;\tau_R) \partial_{\tau_R}G_{\lambda_R}(k_4;\tau_R) G_{\lambda_R}(k_s;\tau_R)\Big]
    &= iP(k_3) P(k_s),
    \\
    \sum_{\lambda_L = \pm} \lambda_L \int_{-\infty}^{\tau_\star} \dd\tau_L\, \partial_{\tau_L}\Big[f_L^2 G_{\lambda_L}(k_1;\tau_L) \partial_{\tau_L}G_{\lambda_L}(k_2;\tau_L) G_{\lambda_L}(k_s;\tau_L)\Big]
    &= iP(k_1)P(k_s),
\end{align}
so that we obtain in total
\begin{align}
    I_2(k_1,k_2;k_3,k_4) = 0,
\end{align}
as expected. 
Once again, we emphasize that this trivially follows from the original expression~\eqref{eq:I2}, and there is no need to perform the derivatives explicitly, as we have done here, in the actual calculations;
we have done so just for comparison.

\subsubsection*{One derivative on external propagators}

Next let us consider the contribution with one time derivative acting on the external propagators and the other acting on the internal propagator.
We denote such a contribution as
\begin{align}
    \left.\big\langle \hat{\phi}_{k_1}(\tau_\star) \hat{\phi}_{k_2}(\tau_\star) \hat{\phi}_{k_3}(\tau_\star) \hat{\phi}_{k_4}(\tau_\star)\big\rangle'\right\vert_{1e}
    &= I_{1}(k_1,k_2;k_3,k_4) + (\mathrm{perms.}),
\end{align}
where
\begin{align}
	I_1(&k_1,k_2;k_3,k_4) = -4\int_C\dd\tau_L \int_C\dd\tau_R\, \partial_{\tau_L}\partial_{\tau_R}\Big[f_L^2 f_R^2 
	\nonumber \\
	&\times G_C(k_1;\tau_L, \tau_\star)  G_C(k_2;\tau_L,\tau_\star) G_C(k_3;\tau_R,\tau_\star) \partial_{\tau_R}G_C(k_4;\tau_R,\tau_\star)
	\partial_{\tau_L}G_C(k_s;\tau_L,\tau_R)
	\Big].
    \label{eq:I1}
\end{align}
Again, this reduces to the boundary contributions at $\tau_{L,R} \to -\infty\pm i0^+$ and thus vanishes.

Let us take the now-familiar procedure of taking the derivatives explicitly.
For the purpose of comparison, it is useful to decompose the internal propagator as
\begin{align}
    G_C(k_s;\tau_L,\tau_R) = \sum_{\lambda_L,\lambda_R=\pm} \theta_C(\lambda_L(\tau_\star-\tau_L)) \theta_C(\lambda_R(\tau_\star-\tau_R)) G_{\lambda_L\lambda_R}(k_s;\tau_L,\tau_R),
\end{align}
so that we can extract the delta functions due to the $\TCstar$-product beyond those arising from the time derivatives acting on $G_{++}(k_s;\tau_L,\tau_R)$ and $G_{--}(k_s;\tau_L,\tau_R)$; the latter are included in the previous literature while the former are not. 
Since $\sum_{\lambda_L = \pm} \lambda_L G_{\lambda_L\lambda_R}(k_s;\tau_\star,\tau_R) = 0$, we then obtain
\begin{align}
    \partial_{\tau_L}G_C(k_s;\tau_L,\tau_R) = \sum_{\lambda_L,\lambda_R=\pm} \theta_C(\lambda_L(\tau_\star-\tau_L)) \theta_C(\lambda_R(\tau_\star-\tau_R)) \partial_{\tau_L}G_{\lambda_L\lambda_R}(k_s;\tau_L,\tau_R).
\end{align}
With this in mind, the time derivatives can be explicitly performed as
\begin{align}
	I_1(&k_1,k_2;k_3,k_4) = -4\sum_{\lambda_L,\lambda_R = \pm} \lambda_L \lambda_R\int_{-\infty}^{\tau_\star} \dd\tau_L \int_{-\infty}^{\tau_\star} \dd\tau_R\, \partial_{\tau_L}\partial_{\tau_R}
	\nonumber \\
	&\times\Big[f_L^2 f_R^2 G_{\lambda_L}(k_1;\tau_L)  G_{\lambda_L}(k_2;\tau_L) G_{\lambda_R}(k_3;\tau_R) \partial_{\tau_R}G_{\lambda_R}(k_4;\tau_R)
	\partial_{\tau_L}G_{\lambda_L\lambda_R}(k_s;\tau_L,\tau_R)
	\Big]
    \nonumber \\
    &+4 f^2(\tau_\star) P(k_1)P(k_2) \sum_{\lambda_L,\lambda_R=\pm} \lambda_L \lambda_R\int_{-\infty}^{\tau_\star}\dd\tau_R\,\partial_{\tau_R}
    \nonumber \\
    &\times
    \Big[f_R^2 G_{\lambda_R}(k_3;\tau_R) \partial_{\tau_R}G_{\lambda_R}(k_4;\tau_R)
	\big[\partial_{\tau_L}G_{\lambda_L\lambda_R}(k_s;\tau_L,\tau_R)\big]_{\tau_L\to\tau_\star}\Big]
    \nonumber \\
    &+4iP(k_3)\sum_{\lambda_L=\pm}\lambda_L\int_{-\infty}^{\tau_\star} \dd\tau_L\,\partial_{\tau_L}\Big[f_L^2 G_{\lambda_L}(k_1;\tau_L)  G_{\lambda_L}(k_2;\tau_L) \partial_{\tau_L}G_{\lambda_L}(k_s;\tau_L)
    \Big]
    \nonumber \\
    &+4 P(k_1) P(k_2) P(k_3).
    \label{eq:I1_split}
\end{align}
As before, the term in the first two lines is the naive contribution, whereas the remaining terms arise from the delta functions associated with the $\TCstar$-product.
Although the first term does not vanish by itself, in total we obtain
\begin{align}
    I_1(k_1,k_2;k_3,k_4) = 0,
\end{align}
confirming our expectation once again.

\subsection{Comment on previous literature}

Several references~\cite{Braglia:2024zsl,Wang:2025qfh,Ma:2026gdh} have recently claimed that the total time derivative term contributes nontrivially to the in-in correlators based on the path integral formalism.\footnote{
    See also \cite{Arroja:2011yj,Burrage:2011hd,Rigopoulos:2011eq,Garcia-Saenz:2019njm,Sou:2022nsd,Ning:2023ybc} for earlier discussions using different methods including path integral formalism. Since we have already  clarified cancellation of the total time derivatives, we do not comment on each paper separately.
}
However, these calculations do not properly implement the $\TCstar$-ordering and consequently miss the contact terms arising from derivatives of the contour-ordering step functions at the turning point.
Once again, our calculations in the in-in formalism have one-to-one correspondence with those in the in-out formalism.
Therefore, omitting the $\TCstar$-product delta functions is equivalent to omitting the corresponding $\Tstar$-product delta functions in the in-out formalism (see Sec.~\ref{subsec:3pt_explicit}), contradicting the well-established fact that the total derivative terms do not perturbatively affect the correlators in the in-out formalism.

On the other hand,~\cite{Kawaguchi:2024lsw} studied the total time derivative term in the path integral and concluded that it does not contribute to the in-in correlators.
Although we agree with their conclusion, our approach differs: 
Ref.~\cite{Kawaguchi:2024lsw} extended the turning point of the in-in contour $\tau_f$ beyond the evaluation point $\tau_\star$, mentioning that the cancellation seems not to occur for $\tau_f = \tau_\star$.
For $\tau_f > \tau_\star$, the inserted operators lie on $C_+$ (or $C_-$) and are dealt with by the standard $\mathcal{T}^\star$-product (or the $\bar{\mathcal{T}}^\star$-product), which seems to motivate their choice $\tau_f > \tau_\star$.
Although there is no issue with taking $\tau_f > \tau_\star$, our point is that this is not necessary; by introducing the contour ordering and the $\TCstar$-product, we can take $\tau_f = \tau_\star$ without any issue.
We believe that the apparent difficulty they encountered for $\tau_f=\tau_\star$ arises from not accounting for the $\TCstar$-product and the associated contact terms.

\section{Summary}
\label{sec:discussion_and_conclusions}

In this paper we showed that, once the contact terms associated with the $\TCstar$-product are taken into account, total time derivative terms do not contribute to in-in correlators at the observation time $\tau = \tau_\star$ even when the quantity inside the total derivative depends on time derivatives of the fields.
In the in-in formalism, the turning point $\tau = \tau_\star$ lies in the interior of a single smooth path integral contour and cannot generate a boundary contribution.
This is for the same reason that an artificial split of the in-out contour cannot affect its correlators.

We demonstrated this result starting from canonical quantization of a system with a total time derivative term in Sec.~\ref{sec:canonical-quantization}. 
Despite the appearance of the higher-derivative term, the second-class constraints remove the additional variables, recovering the dynamics of the system without the total derivative.
We then showed in Secs.~\ref{sec:path-integral-tstar} and~\ref{sec:closed-time-path} that, in the discretized path integral, the total derivative can be reintroduced by a finite difference whose sum trivially vanishes. 
The continuum limit of the path integral automatically leads to the $\TCstar$-product in the in-in formalism, with derivatives acting outside contour ordering. 
The $\TCstar$-product generates additional contact terms that were overlooked in the previous studies.
Our explicit three- and four-point calculations in Sec.~\ref{sec:examples} demonstrate that these contact terms cancel the apparent total derivative contributions at $\tau = \tau_\star$ reported in these studies.

Rewriting the action in equivalent but more convenient forms is a standard tool often used in quantum field theory.
Our analysis confirms that adding or removing total time derivatives provides one such tool for calculating cosmological correlators.
With the $\TCstar$-product properly accounted for, these terms generate no boundary contributions at the observation time $\tau = \tau_\star$ and leave the correlators unchanged.

\section*{Acknowledgment}

This work is supported in part by JSPS KAKENHI Grant Numbers JP24K07013~(R.J.) and JP26K07096~(K.M.).

\appendix
\section{Hamilton's equations}
\label{app:Hamiltonian_evolution}

In this Appendix, we derive Hamilton's equations when explicit time-dependence is present.

We begin with the cases without explicit time-dependence.
We introduce a symplectic potential $\Theta$, whose components are $\Theta_a(z)$, so that
\begin{align}
    \Theta = \Theta_a(z)\dd z^a,
    \label{eq:symplectic_potential}
\end{align}
where $z^a$ are coordinates of the phase space.
The first-order phase-space action is
\begin{align}
    S[z] = \int^{t_\mathrm{f}}_{t_\mathrm{i}} \dd t\, \left[\Theta_a(z)\dot{z}^a-H(z)\right],
    \label{eq:app_ps_action}
\end{align}
where $H(z)$ is the Hamiltonian function.
Together with the symplectic form, it determines the time evolution of the system.
Varying the action and integrating by parts gives
\begin{align}
    \delta S=\int^{t_\mathrm{f}}_{t_\mathrm{i}} \dd t\, \left[
    \left( \frac{\partial \Theta_a}{\partial z^b} \dot{z}^a - \frac{\partial \Theta_b}{\partial z^a} \dot{z}^a\right) \delta z^b-\frac{\partial H}{\partial z^b}\delta z^b\right]
    +\left[\Theta_a \delta z^a \right]^{t_\mathrm{f}}_{t_\mathrm{i}}.
    \label{eq:delta_S_1order}
\end{align}
The bulk term gives Hamilton's equations:
\begin{align}
    \left( \frac{\partial \Theta_a}{\partial z^b}  - \frac{\partial \Theta_b}{\partial z^a} \right)\dot{z}^a=\frac{\partial H}{\partial z^b}.
    \label{eq:eom_H}
\end{align}
The boundary contribution $\left[\Theta_a \delta z^a \right]^{t_\mathrm{f}}_{t_\mathrm{i}}$ vanishes as long as the initial and final points are kept fixed.
One example is a system with a single canonical degree of freedom $\Theta=p \dd q$.
From Eq.~\eqref{eq:eom_H}, we can write down the familiar Hamilton's equations:
\begin{align}
    \dot{q}=\frac{\partial H}{\partial p}, \qquad
    \dot{p}=-\frac{\partial H}{\partial q}.
\end{align}

We can generalize the above discussion to the cases where $\Theta_a$ and $H$ have explicit time dependence.
Instead of Eq.~\eqref{eq:delta_S_1order}, we have
\begin{align}
    \delta S=\int^{t_\mathrm{f}}_{t_\mathrm{i}} \dd t\, \left[
    \left( \frac{\partial \Theta_a}{\partial z^b} \dot{z}^a - \frac{\partial \Theta_b}{\partial z^a} \dot{z}^a- \partial_t \Theta_b\right) \delta z^b-\frac{\partial H}{\partial z^b}\delta z^b\right]
    +\left[\Theta_a \delta z^a \right]^{t_\mathrm{f}}_{t_\mathrm{i}},
\end{align}
and obtain Hamilton's equations as
\begin{align}
    \left( \frac{\partial \Theta_a}{\partial z^b}  - \frac{\partial \Theta_b}{\partial z^a} \right)\dot{z}^a=\frac{\partial H}{\partial z^b} + \partial_t \Theta_b.
\end{align}
We consider an example related to the main text: one canonical degree of freedom with the symplectic potential
\begin{align}
    \Theta=p \dd q + \dd_{\rm ph} G
    = p \dd q + \frac{\partial G}{\partial z^a} \dd z^a,
\end{align}
where $\dd_{\rm ph}$ denotes the exterior derivative on the phase space at fixed time.
Hamilton's equations are then
\begin{align}
    \dot{q}=\frac{\partial H}{\partial p}+\frac{\partial^2 G}{\partial t \partial p}, \qquad
    \dot{p}=-\frac{\partial H}{\partial q} - \frac{\partial^2 G}{\partial t \partial q}.
    \label{eq:app_EoM_G}
\end{align}
Identifying $H$ with $H_{\rm red}$ in the main text, we see that the partial derivatives of $G$ in these equations cancel with the contribution from $G_t$ in $H_{\rm red}$, and therefore we have Hamilton's equations for a canonical field as expected.

\bibliographystyle{utphys}
\bibliography{ref}

@article{Kawaguchi:2024lsw,
    author = "Kawaguchi, Ryodai and Tsujikawa, Shinji and Yamada, Yusuke",
    title = "{Roles of boundary and equation-of-motion terms in cosmological correlation functions}",
    eprint = "2403.16022",
    archivePrefix = "arXiv",
    primaryClass = "hep-th",
    reportNumber = "WUCG-24-03",
    doi = "10.1016/j.physletb.2024.138962",
    journal = "Phys. Lett. B",
    volume = "856",
    pages = "138962",
    year = "2024"
}

@article{Ma:2026gdh,
    author = "Ma, Yanjiao and Wang, Dong-Gang and Wang, Xiangwei and Wang, Yi and Yu, Wenqi",
    title = "{On Cosmological Correlators with Boundary Contributions}",
    eprint = "2606.06282",
    archivePrefix = "arXiv",
    primaryClass = "gr-qc",
    month = "6",
    year = "2026"
}

@article{Wang:2025qfh,
    author = "Wang, Dong-Gang and Wang, Xiangwei and Wang, Yi and Yu, Wenqi",
    title = "{Interact or Twist: Cosmological Correlators from Field Redefinitions Revisited}",
    eprint = "2508.12856",
    archivePrefix = "arXiv",
    primaryClass = "hep-th",
    month = "8",
    year = "2025"
}

@article{Landsman:1986uw,
	author = {Landsman, N. P. and van Weert, C. G.},
	doi = {10.1016/0370-1573(87)90121-9},
	journal = {Phys. Rept.},
	pages = {141--249},
	reportnumber = {ITFA-86-12},
	title = {{Real and Imaginary Time Field Theory at Finite Temperature and Density}},
	volume = {145},
	year = {1987}}

@article{Braglia:2024zsl,
	archiveprefix = {arXiv},
	author = {Braglia, Matteo and Pinol, Lucas},
	doi = {10.1007/JHEP08(2024)068},
	eprint = {2403.14558},
	journal = {JHEP},
	pages = {068},
	primaryclass = {astro-ph.CO},
	title = {{No time to derive: unraveling total time derivatives in in-in perturbation theory}},
	volume = {08},
	year = {2024}}

@article{Rigopoulos:2011eq,
    author = "Rigopoulos, Gerasimos",
    title = "{Gauge invariance and non-Gaussianity in Inflation}",
    eprint = "1104.0292",
    archivePrefix = "arXiv",
    primaryClass = "astro-ph.CO",
    reportNumber = "TTK-11-11",
    doi = "10.1103/PhysRevD.84.021301",
    journal = "Phys. Rev. D",
    volume = "84",
    pages = "021301",
    year = "2011"
}

@article{Ning:2023ybc,
    author = "Ning, Sirui and Sou, Chon Man and Wang, Yi",
    title = "{On the decoherence of primordial gravitons}",
    eprint = "2305.08071",
    archivePrefix = "arXiv",
    primaryClass = "hep-th",
    doi = "10.1007/JHEP06(2023)101",
    journal = "JHEP",
    volume = "06",
    pages = "101",
    year = "2023"
}

@article{Sou:2022nsd,
    author = "Sou, Chon Man and Tran, Duc Huy and Wang, Yi",
    title = "{Decoherence of cosmological perturbations from boundary terms and the non-classicality of gravity}",
    eprint = "2207.04435",
    archivePrefix = "arXiv",
    primaryClass = "hep-th",
    doi = "10.1007/JHEP04(2023)092",
    journal = "JHEP",
    volume = "04",
    pages = "092",
    year = "2023"
}

@article{Garcia-Saenz:2019njm,
    author = "Garcia-Saenz, Sebastian and Pinol, Lucas and Renaux-Petel, S{\'e}bastien",
    title = "{Revisiting non-Gaussianity in multifield inflation with curved field space}",
    eprint = "1907.10403",
    archivePrefix = "arXiv",
    primaryClass = "hep-th",
    doi = "10.1007/JHEP01(2020)073",
    journal = "JHEP",
    volume = "01",
    pages = "073",
    year = "2020"
}

@article{Burrage:2011hd,
    author = "Burrage, Clare and Ribeiro, Raquel H. and Seery, David",
    title = "{Large slow-roll corrections to the bispectrum of noncanonical inflation}",
    eprint = "1103.4126",
    archivePrefix = "arXiv",
    primaryClass = "astro-ph.CO",
    reportNumber = "DAMTP-2011-19",
    doi = "10.1088/1475-7516/2011/07/032",
    journal = "JCAP",
    volume = "07",
    pages = "032",
    year = "2011"
}

@article{Arroja:2011yj,
    author = "Arroja, Frederico and Tanaka, Takahiro",
    title = "{A note on the role of the boundary terms for the non-Gaussianity in general k-inflation}",
    eprint = "1103.1102",
    archivePrefix = "arXiv",
    primaryClass = "astro-ph.CO",
    doi = "10.1088/1475-7516/2011/05/005",
    journal = "JCAP",
    volume = "05",
    pages = "005",
    year = "2011"
}

@article{Weinberg:2005vy,
    author = "Weinberg, Steven",
    title = "{Quantum contributions to cosmological correlations}",
    eprint = "hep-th/0506236",
    archivePrefix = "arXiv",
    reportNumber = "UTTG-01-05",
    doi = "10.1103/PhysRevD.72.043514",
    journal = "Phys. Rev. D",
    volume = "72",
    pages = "043514",
    year = "2005"
}

@article{Arzt:1993gz,
    author = "Arzt, Christopher",
    title = "{Reduced effective Lagrangians}",
    eprint = "hep-ph/9304230",
    archivePrefix = "arXiv",
    reportNumber = "UM-TH-92-28",
    doi = "10.1016/0370-2693(94)01419-D",
    journal = "Phys. Lett. B",
    volume = "342",
    pages = "189--195",
    year = "1995"
}

@article{Brivio:2017vri,
    author = "Brivio, Ilaria and Trott, Michael",
    title = "{The Standard Model as an Effective Field Theory}",
    eprint = "1706.08945",
    archivePrefix = "arXiv",
    primaryClass = "hep-ph",
    doi = "10.1016/j.physrep.2018.11.002",
    journal = "Phys. Rept.",
    volume = "793",
    pages = "1--98",
    year = "2019"
}

@article{Burgess:2007pt,
    author = "Burgess, C. P.",
    title = "{Introduction to Effective Field Theory}",
    eprint = "hep-th/0701053",
    archivePrefix = "arXiv",
    doi = "10.1146/annurev.nucl.56.080805.140508",
    journal = "Ann. Rev. Nucl. Part. Sci.",
    volume = "57",
    pages = "329--362",
    year = "2007"
}

@article{Manohar:2018aog,
    author = "Manohar, Aneesh V.",
    editor = "Davidson, Sacha and Gambino, Paolo and Laine, Mikko and Neubert, Matthias and Salomon, Christophe",
    title = "{Introduction to Effective Field Theories}",
    eprint = "1804.05863",
    archivePrefix = "arXiv",
    primaryClass = "hep-ph",
    doi = "10.1093/oso/9780198855743.003.0002",
    month = "4",
    year = "2018"
}

@book{Dirac1964,
	address = {New York},
	author = {Dirac, P. A. M.},
	number = {2},
	publisher = {Belfer Graduate School of Science, Yeshiva University},
	series = {Belfer Graduate School of Science Monographs Series},
	title = {Lectures on Quantum Mechanics},
	year = {1964}}

@article{Liu:2018kfw,
    author = "Liu, Hong and Glorioso, Paolo",
    title = "{Lectures on non-equilibrium effective field theories and fluctuating hydrodynamics}",
    eprint = "1805.09331",
    archivePrefix = "arXiv",
    primaryClass = "hep-th",
    reportNumber = "MIT-CTP/5018; EFI-18-8, MIT-CTP-5018, EFI-18-8",
    doi = "10.22323/1.305.0008",
    journal = "PoS",
    volume = "TASI2017",
    pages = "008",
    year = "2018"
}

@article{Haehl:2024pqu,
    author = "Haehl, Felix M. and Rangamani, Mukund",
    title = "{Records from the S-Matrix Marathon: Schwinger-Keldysh Formalism}",
    eprint = "2410.10602",
    archivePrefix = "arXiv",
    primaryClass = "hep-th",
    month = "10",
    year = "2024"
}

@article{Berges:2004yj,
    author = "Berges, Juergen",
    editor = "Bracco, Mirian and Chiapparini, Marcelo and Ferreira, Erasmo and Kodama, Takeshi",
    title = "{Introduction to nonequilibrium quantum field theory}",
    eprint = "hep-ph/0409233",
    archivePrefix = "arXiv",
    doi = "10.1063/1.1843591",
    journal = "AIP Conf. Proc.",
    volume = "739",
    number = "1",
    pages = "3--62",
    year = "2004"
}

@article{Haehl:2016pec,
    author = "Haehl, Felix M. and Loganayagam, R. and Rangamani, Mukund",
    title = "{Schwinger-Keldysh formalism. Part I: BRST symmetries and superspace}",
    eprint = "1610.01940",
    archivePrefix = "arXiv",
    primaryClass = "hep-th",
    doi = "10.1007/JHEP06(2017)069",
    journal = "JHEP",
    volume = "06",
    pages = "069",
    year = "2017"
}

@book{Bastianelli:2006rx,
    author = "Bastianelli, F. and van Nieuwenhuizen, P.",
    title = "{Path integrals and anomalies in curved space}",
    doi = "10.1017/CBO9780511535031",
    isbn = "978-0-511-21772-2, 978-0-521-12050-0, 978-0-521-84761-2",
    publisher = "Cambridge University Press",
    series = "Cambridge Monographs on Mathematical Physics",
    month = "9",
    year = "2006"
}

@book{Srednicki:2007qs,
    author = "Srednicki, M.",
    title = "{Quantum field theory}",
    doi = "10.1017/CBO9780511813917",
    isbn = "978-0-521-86449-7, 978-0-511-26720-8",
    publisher = "Cambridge University Press",
    month = "1",
    year = "2007"
}

@book{Weinberg:1995mt,
    author = "Weinberg, Steven",
    title = "{The Quantum theory of fields. Vol. 1: Foundations}",
    doi = "10.1017/CBO9781139644167",
    isbn = "978-0-521-67053-1, 978-0-511-25204-4",
    publisher = "Cambridge University Press",
    month = "6",
    year = "2005"
}

@book{Itzykson:1980rh,
    author = "Itzykson, C. and Zuber, J. B.",
    title = "{Quantum Field Theory}",
    publisher = "McGraw-Hill",
    address = "New York",
    series = "International Series In Pure and Applied Physics",
    year = "1980"
}

@article{Gross:1969zow,
    author = "Gross, D. J. and Jackiw, R.",
    title = "{Construction of covariant and gauge invariant {$T^\star$} products}",
    doi = "10.1016/0550-3213(69)90207-7",
    journal = "Nucl. Phys. B",
    volume = "14",
    pages = "269--284",
    year = "1969"
}

@book{HenneauxTeitelboim1992,
	address = {Princeton},
	author = {Henneaux, Marc and Teitelboim, Claudio},
	isbn = {9780691037691},
	publisher = {Princeton University Press},
	title = {Quantization of Gauge Systems},
	year = {1992}}

@article{Maldacena:2002vr,
    author = "Maldacena, Juan Martin",
    title = "{Non-Gaussian features of primordial fluctuations in single field inflationary models}",
    eprint = "astro-ph/0210603",
    archivePrefix = "arXiv",
    doi = "10.1088/1126-6708/2003/05/013",
    journal = "JHEP",
    volume = "05",
    pages = "013",
    year = "2003"
}

@book{Kugo:1997gva,
    author = "Kugo, Taichiro",
    title = "{Eichtheorie}",
    doi = "10.1007/978-3-642-59128-0",
    isbn = "978-3-540-62063-1, 978-3-642-63860-2, 978-3-642-59128-0",
    publisher = "Springer",
    year = "1997"
}

\end{document}